\documentclass[traditabstract]{aa} 
\usepackage{graphicx,natbib,longtable,txfonts,lscape}
\usepackage[switch]{lineno}

\newcommand{\kms}{$\rm{\,km \,s}^{-1}$}

\newcommand{\WHz}{\>{\rm W}\,{\rm Hz}^{-1}}

\usepackage{color}

\begin{document}

\title{The radio properties of $z>3.5$ quasars:\\ Are most high-redshift radio-loud active galactic nuclei obscured? } \author{Alessandro
  Capetti\inst{1} \and Barbara Balmaverde\inst{1}}
\titlerunning{The radio properties of $z>3.5$ quasars}

\institute{INAF - Osservatorio Astrofisico di Torino, Strada
  Osservatorio 20, I-10025 Pino Torinese, Italy } \date{}

\abstract{We explore the radio properties of powerful (rest-frame
  luminosity $\gtrsim10^{28} \WHz$ at 500 MHz) high-redshift ($z
  \gtrsim 3.5$) quasars. The aim of this study is to gain a better
  understanding of radio-loud sources at the epoch when they reach the
  highest space density. We selected 29 radio-loud quasars at low
  radio frequencies (76 MHz). Their radio spectra, covering the range
  from 76 MHz to 5 GHz, are generally well reproduced by a single
  power law. We created samples that were matched in radio luminosity at lower
  redshift (from $z\sim1.3$ to $z\sim2.8$) to investigate any spectral
  evolution. We find that the fraction of flat-spectrum radio quasars
  (FSRQs) strongly increases with redshift (from $\sim 8\%$ at z=1.2
  to $\sim 45\%$ at z$>$3.5).  This effect is also observed in quasars with
  lower luminosities (down to $\sim10^{27} \WHz$). The increase in the
  fraction of FSRQs with redshift corresponds to a decrease in the steep-  spectrum radio quasars.

  This result can be explained when we assume that the beaming factor
    and the slope of the luminosity function do not change with
    redshift and if high-redshift radio-loud sources can be recognized
  as quasars only when they are seen at a small viewing angle
  ($\lesssim 25^\circ$), but most of them, about 90\%, are
  obscured in the UV and optical bands.  We also found a trend for the
  size of radio sources to decrease with increasing redshift. Because
  projection effects are insufficient to cause this trend, we suggest
  that the large amount of gas causing the nuclear obscuration also
  hampers the growth of the more distant sources.}

\maketitle

\section{Introduction}
\label{intro}

Powerful radio-loud active galactic nuclei (RLAGN) represent the most
extreme manifestation of activity powered by accretion onto
supermassive black holes. They play a pivotal role in cosmology and
galaxy evolution. Their host galaxies are, or are destined to become,
massive giant elliptical galaxies, many of which are brightest cluster
galaxies (see, e.g., \citealt{best05b,miraghaei17}). The intense
nuclear emission and highly collimated jets of RLAGN can profoundly
influence the star formation, excitation, and ionization of the
intra-cluster medium (ICM; e.g.,
\citealt{fabian12,voit15}). Cosmological models that do not account
for the radio mode AGN feedback lead to galaxy luminosities, colors,
and disk and bulge sizes that disagree with observations (e.g.,
\citealt{croton06}).

At low redshift, $z \lesssim 1$, the properties of radio sources are
quite well understood (see \citealt{tadhunter16} for a review).  In
particular, it has been suggested that the observed variegated
phenomenology can be understood within the unified model (UM) of radio
AGN. The UM postulates that various classes of objects are
intrinsically identical and differ only in their orientation with
respect to our line of sight (see, e.g., \citealt{antonucci93} for a
review). The origin of the observational differences is due to the
presence of i) circumnuclear absorbing material (usually referred to
as the obscuring torus) that produces selective absorption when the
source is observed at a large angle from its radio axis, and ii) Doppler
boosting, which is associated with relativistic motions in AGN jets.

\begin{figure*}
    \includegraphics[width=0.33\textwidth]{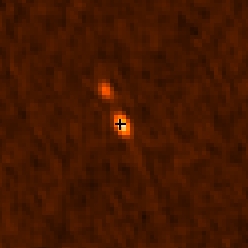}
    \includegraphics[width=0.33\textwidth]{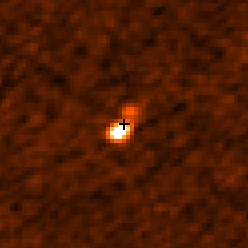}
    \includegraphics[width=0.33\textwidth]{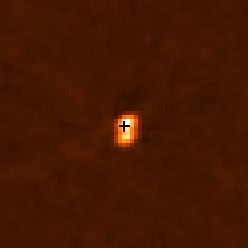}
    \caption{VLASS images at 2\farcs5 resolution of the three
      HzRQs associated with extended radio sources. From left to
      right: SDSS~034402.85-065300.6 (at z=3.942),
      SDSS~085023.56+153003.9 (at z=3.713), and
      SDSS~145805.22+172524.5 (at z=3.657). The field of view is
      60\arcsec$\times$60\arcsec, corresponding to $\sim$ 430 kpc
        $\times$ 430 kpc. The black cross marks the optical location
      of the QSO.}
    \label{vlass}
\end{figure*}

In the unification scheme of RLAGN (e.g., \citealt{urry95})
narrow-lined radio galaxies of Fanaroff Riley (FR) type II 
\citep{fanaroff74} and broad-lined FR~IIs together with RL QSOs are
considered to be intrinsically indistinguishable.  Their different
aspect (in particular, the absence of broad emission lines in FR~IIs)
is only related to their orientation in the sky with respect to our
line of sight. Therefore, in its stricter interpretation of a pure
orientation scheme, the UM predicts that narrow- and broad-line FR~II
are drawn from the same parent population. According to this model, a
source appears as a quasar stellar object (QSO) only when its radio
axis is oriented within a small cone from the observer's line of sight
\citep{barthel89}; conversely, when seen at large angles, the nuclear
emission (including the broad-line region) is obscured, and they are
called radio galaxies. Probably the most convincing of several pieces
of evidence in favor of the UM is the detection of broad lines in the
polarized spectra of narrow-line objects
\citep{antonucci82,antonucci84}, which was interpreted as the result
of scattered light from an otherwise obscured nucleus.  Although the
UM is generally successful in explaining the variety of RLAGN
behavior, some issues still remain to be explained. For example, the
powerful FR~II showing weak emission lines might represent evidence
that in addition to the effect of orientation, changes in the
accretion rate play an important role (see, e.g.,
\citealt{tadhunter98}).

\citet{baldi13} explored the properties of radio sources extracted
from the third Cambridge catalog (3C, \citealt{spinrad85}) with
$z<0.3$ and found a fraction of narrow-line sources of $\sim 65$\%. In
a torus geometry, this fraction corresponds to a torus opening angle
of $50^{\circ} \pm 5$.  \citet{lawrence91}, \citet{hill96}, and
\citet{willott00} found that the fraction of narrow-line objects
decreases with increasing radio power. It reaches $\sim 50\%$ for the
most powerful sources at $1<z<2$. The authors proposed that this is
due to a lower covering factor of the circumnuclear absorption
structure in the more luminous sources, the so-called receding torus
model.

The UM has also successfully explained the observed properties of the
sources in which the dominant effects of anisotropy are due to Doppler
amplification of the emission of a relativistic jet oriented at a
small angle with respect to our line of sight (see \citealt{urry95}
for a review). The highly amplified sources that form the class of
blazars can be separated into two main classes depending on whether broad emission lines are visible in their optical spectra. In
the first case, they are called flat-spectrum radio quasars (FSRQs; with
a radio spectral index $\alpha < 0.5$) and their observed radio
emission originates mainly from the base of their relativistic jets
when these are seen at a small angle from the line of sight. In
  the objects oriented at a larger angle, in which the amplification
  is not so extreme, the radio emission is ascribed to optically thin
  synchrotron radiation produced by large-scale regions. Therefore,
  they show a steep radio spectrum and are called steep-spectrum radio
  quasars (SSRQs).

The high-redshift range, $z \gtrsim 3 - 4 $, is particularly
important: The results of \citet{caccianiga19} and \citet{lister19}
showed that the density of the most luminous RLAGN peaks at this epoch,
when AGN feedback was more active. As already observed at low
redshift, high-z RLAGN might also manifest as sources in which the
nucleus is optically obscured, as in the case of high-redshift radio
galaxies (HzRGs). The nucleus can instead be directly visible, as
  is the case in high-redshift radio quasars (HzRQs). HzRQs in turn might show
both steep and flat spectra (e.g., \citealt{Sotnikova21}), suggesting
that the RLAGN unified scheme might also apply at high redshift.

  However, our knowledge of the properties of RLAGN at high redshift
  is less detailed than it is for lower redshift objects.  Only 19 HzRGs are contained in the list of optically obscured AGN at
  $z>3.5$ in the list compiled by \citet{miley08}, and only a few more
  have been discovered since then (e.g.,
  \citealt{saxena18,yamashita20}). Furthermore, the main programs of
  HzRG searches were carried out selecting ultrasteep radio
  sources (USS; e.g., \citealt{debreuck06}). USS were defined by
  \citet{miley89} as sources with a spectral index $\alpha > 1$ (defined
  as $f_\nu \propto\nu^{-\alpha}$). This criterion was motivated by
  the phenomenological trend that suggested that higher-redshift radio
  sources have steeper radio spectra \citep{debreuck01}, but it leads
  to a selection bias.

  Conversely, large samples of quasars at high redshift are available.
  For example, there are $\sim 10,000$ QSOs at z$>$3.5 in the
  Sloan Digital Sky Survey (SDSS; \citealt{york00}), and a significant
  fraction of them ($\sim 8$\%) are
  radio-loud, regardless of redshift \citep{ivezic02}. The UM suggests that the study of HzRQs
  can be used to explore the nature, evolution, and properties of the
  general population of powerful radio AGN in the early Universe,
  overcoming the problem of the paucity of known HzRGs.

  Several studies have explored the radio properties of HzRQs. For
    example, \citet{coppejans17} found that in 45 RQs with
    $z>4.5$, flat, steep, and peaked spectra are present in almost
    equal number. From high-resolution observations,
    \citet{krezinger22} found that about half of these are highly core
    dominated blazars, while the remaining sources are misaligned
    AGN, that is, SSRQs.

However, these works were performed based on observations in the gigahertz
regime, which is a strong limitation, particularly for high-redshift
  objects. To gain better insight into the properties of HzRQs, it is
very important to access the low radio frequency window down to
  $\sim 100$ MHz, where the steep-spectrum radio emission from the
  extended structures is more prominent and can be studied in greater
  detail.  Furthermore, low-frequency observations enlarge the
coverage of the radio spectra, which is necessary to investigate the overall
behavior of the radio emission. Only a few studies took
advantage of low-frequency data to date. \citet{gurkan19} used data from the
150 MHz Low-Frequency ARray (LOFAR; \citealt{vanhaarlem13}) to
investigate the behavior of the radio-loudness parameter based on low-frequency observations. \citet{Gloudemans21} analyzed radio QSOs at
$z>5$ and found that 36\% were detected by LOFAR. The median
radio spectral index from 150 MHz to 1.4 GHz index is
$\alpha=0.29$. \citet{gloudemans23} instead built radio spectra of RQs
with $z>5.6$, finding no sign of a spectral flattening between 144 MHz
and a few gigahertz, indicating that there is no strong absorption in the radio
band. Nonetheless, a study of the overall radio properties, including
the low-frequency regime, of a sizeable sample of HzRQs is required to
assess the general behavior of these sources.

Another important issue that needs to be addressed is the connection
between the various classes of high-redshift RLAGN, that is, HzRGs,
SSRQs, and FSRQs, and the orientation and obscuration effects for
their appearance. \citet{volonteri11} first noted that the observed
number of misaligned radio-loud quasars is smaller than expected based
on the number of blazars found at high redshift. To solve this difference,
\citep{ghisellini16} proposed that a large fraction of high-redshift
RLAGN may be obscured. 

In this paper, we focus on the radio properties of powerful HzRQs at
$z>3.5$ derived from broad-band radio spectra and high-resolution
  radio imaging. The paper is organized as follows: In Sect. 2 we
describe the selection of a sample of radio quasars at $z>3.5$, whose
radio properties are presented in Sect. 3. In Sect. 4 we introduce
three samples of lower-redshift RQs that we used to set our findings in a
broader context. We selected only QSOs with the same radio power as
  the sources in the highest-redshift bin, that is, $\gtrsim 10^{28}
  \WHz$ at 500 MHz. In Sect. 5 we extend the analysis to less
powerful radio quasars, $\gtrsim 10^{27} \WHz$. The results are
discussed in Sect. 6, and in Sect. 7 we provide a summary and our
conclusions. We adopt the following set of cosmological parameters:
$H_{\rm 0}=69.7$ \kms\ Mpc$^{-1}$ and $\Omega_m$=0.286
\citep{bennett14}.

\section{Sample selection}
\label{sample}

The sample was selected from the catalog of quasars from the Data
Release 16 (DR16; \citealt{lyke20}) of the SDSS, requiring a redshift
$z>3.5$. We required a detection at 76 MHz by the GLEAM survey
\citep{wayth15}, which is performed with the Murchison Widefield Array (MWA;
\citealt{tingay13}) as compiled in the extragalactic GLEAM catalog
(EGC, \citealt{hurley17}). The catalog covers the area with $\delta <
30^\circ$ at a resolution of $\sim 2^\prime$ with a typical RMS
  noise of $\sim 20$ mJy beam$^{-1}$, and it reaches a completeness
of $\sim 55\%$ at $\sim$ 50 mJy. We preferred the GLEAM catalog to
deeper and higher-resolution surveys, such as that obtained with the Tata
Institute of Fundamental Research Giant Metrewave Radio Telescope
(\citealt{swarup91,intema17}), called Tata Institute of
Fundamental Research GMRT Sky Survey (TGSS), which has a flux density limit of
17.5 mJy, or the 150 MHz the LOFAR Two-metre Sky Survey (LoTSS;
\citealt{shimwell22}), whose flux density limit is $\sim 0.8$
mJy. The reason for this choice is that the EGC provides us
with a broad-band coverage of the low-frequency radio spectrum, with
20 separate measurements in the range from 72 MHz to 231 MHz. This is very
useful for our analysis.

\begin{figure}
    \includegraphics[width=0.48\textwidth]{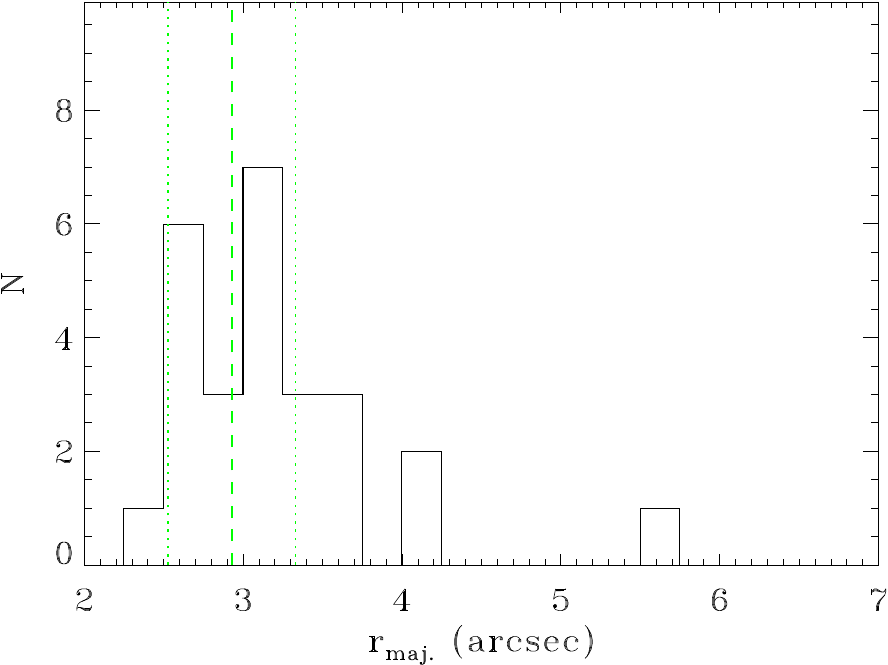}
    \caption{Distribution of the length of the major axis, $r_{\rm
        maj.}$, obtained from a two-dimensional Gaussian fit on the
      VLASS images of the 29 HzRQs. The dashed vertical line marks the
      median value of $r_{\rm maj.}$ of the VLASS sources in the same
      range of declination, and the two dotted lines mark the observed
      spread of $\sim0\farcs4$ observed for $r_{\rm maj.}$.}
    \label{size}
\end{figure}

\begin{table*}[h]
  \caption{Properties of the high-redshift QSO sample.}
    \centering
    \begin{tabular}{l | c r l c}
      Name  & z & $\alpha$ & P$_{{500} {\rm MHz}}$ & r$_{\rm maj.}$\\
      \hline
SDSS~001115.23+144601.8 & 4.964 &  0.43$\pm$0.08 & 28.52 &  5.60$\pm$0.82   \\ 
SDSS~003843.98+031120.9 & 3.674 &  0.73$\pm$0.03 & 29.01 &  3.12$\pm$0.01   \\ 
SDSS~011540.51+035643.3 & 4.189 &  0.58$\pm$0.04 & 29.11 &  3.29$\pm$0.01   \\ 
SDSS~011747.86+011407.6 & 3.694 &  0.77$\pm$0.03 & 28.44 &  3.16$\pm$0.04   \\ 
SDSS~021655.64+051018.3 & 3.817 &  0.45$\pm$0.05 & 28.21 &  3.10$\pm$0.02   \\ 
SDSS~030437.21+004653.5 & 4.266 &  0.37$\pm$0.04 & 28.08 &  3.46$\pm$0.03   \\ 
SDSS~034402.85-065300.6 & 3.942 &  0.77$\pm$0.11 & 28.24 &           E  \\ 
SDSS~075910.06+283310.8 & 3.590 &  0.80$\pm$0.06 & 28.67 &  2.77$\pm$0.01   \\ 
SDSS~080710.74+131739.4 & 3.716 &  0.06$\pm$0.09 & 27.82 &  2.68$\pm$0.01   \\ 
SDSS~082323.32+155206.8 & 3.793 &  0.68$\pm$0.02 & 28.80 &  3.51$\pm$0.02   \\ 
SDSS~083322.49+095941.2 & 3.713 &  0.32$\pm$0.04 & 28.63 &  2.81$\pm$0.01   \\ 
SDSS~083549.42+182520.0 & 4.412 &  0.32$\pm$0.06 & 28.28 &  2.43$\pm$0.01   \\ 
SDSS~085023.56+153003.9 & 3.713 &  0.76$\pm$0.05 & 28.62 &           E  \\ 
SDSS~085257.12+243103.1 & 3.612 &  0.46$\pm$0.06 & 28.73 &  2.56$\pm$0.01   \\ 
SDSS~093113.34+233203.1 & 3.542 &  0.85$\pm$0.05 & 28.60 &  4.19$\pm$0.06   \\ 
SDSS~094004.80+052630.9 & 4.503 &  0.72$\pm$0.04 & 28.97 &  2.58$\pm$0.01   \\ 
SDSS~094948.33+235822.1 & 3.626 &  0.28$\pm$0.09 & 28.24 &  3.51$\pm$0.03   \\ 
SDSS~102107.57+220921.4 & 4.262 &  0.40$\pm$0.04 & 29.01 &  3.17$\pm$0.01   \\ 
SDSS~105407.13-012342.3 & 3.984 &  0.87$\pm$0.03 & 28.94 &  3.08$\pm$0.02   \\ 
SDSS~110147.88+001039.4 & 3.693 &  0.68$\pm$0.03 & 29.28 &  3.39$\pm$0.01   \\ 
SDSS~115339.37+074716.1 & 6.206 &  0.34$\pm$0.06 & 28.97 &  2.61$\pm$0.01   \\ 
SDSS~130906.78+034607.5 & 5.815 &  0.79$\pm$0.03 & 29.60 &  3.58$\pm$0.01   \\ 
SDSS~133605.56+070718.8 & 3.642 &  0.59$\pm$0.06 & 28.56 &  3.13$\pm$0.03   \\ 
SDSS~142048.01+120545.9 & 4.026 &  0.60$\pm$0.03 & 28.92 &  3.11$\pm$0.02   \\ 
SDSS~144516.46+095836.0 & 3.541 & -0.23$\pm$0.07 & 28.82 &  4.24$\pm$0.01   \\ 
SDSS~145805.22+172524.5 & 3.657 &  0.76$\pm$0.02 & 29.14 &           E  \\ 
SDSS~153533.88+025423.3 & 4.388 &  0.44$\pm$0.07 & 28.66 &  2.66$\pm$0.01   \\ 
SDSS~155930.97+030448.1 & 3.895 & -0.12$\pm$0.04 & 28.62 &  2.71$\pm$0.01   \\ 
SDSS~222951.52-012934.4 & 3.504 &  0.51$\pm$0.06 & 28.59 &  2.94$\pm$0.01   \\ 
  \hline
    \end{tabular}
    
    \smallskip
    Column description: (1) SDSS name, (2) redshift, (3) radio
    spectral index with the spectrum reproduced as $F_{\nu} \propto
    \nu^{-\alpha}$, (4) rest-frame radio luminosity at 500 MHz in
    $\WHz$ derived from the fitted radio spectra, (5) measurement of
    the major axis for the compact radio sources in arcsec. An 'E'
    marks an extended radio sources.
    \label{tab1}
\end{table*}

We adopted a search radius for the association between quasars and
GLEAM sources of 30$\arcsec$, yielding 46 sources. The genuine
radio-optical association can be verified with the radio
images of the \textit{Karl G. Jansky} Very Large Array Sky Survey (VLASS;
\citealt{lacy20}), which are obtained at a frequency of 3 GHz with a spatial
resolution of 2\farcs5. This analysis shows in 7 cases that the
QSO is not detected by the VLASS, and the low-frequency radio
flux seen by GLEAM is due to a nearby source. In 5 further cases,
the QSO emits in the radio, but a brighter radio source is found
nearby. The inspection of the TGSS 150 MHz images confirms that this
is the dominant source at low radio frequencies as well. With this
procedure, we discarded these 12 sources from the initial
sample. Finally, we visually inspected the spectra of the selected
objects and found that the redshift reported in the DR16 catalog for 5 of them is
incorrect. We therefore focused on the remaining 29 quasars.

The main properties of the selected QSO are tabulated in
Table \ref{tab1}. The redshift of the sources ranges from 3.5 to 6.5,
with a median of $z = 3.8$. We estimated the radio-loudness parameter, which is
defined following \citet{jiang07} as $R=f_{\rm 6 cm}/f_{2500}$, where
$f_{\rm 6 cm}$ and $f_{2500}$ are the flux densities at a rest frame of 6
cm and 2500 \AA, respectively. We complemented the values of $R$
that were tabulated for most sources by \citet{shen11} with our own
measurements. We find that all sources have very high values of $R$,
ranging from $10^2$ to 10$^4$, which are to be compared to the standard
radio-loudness threshold of $R \sim 10$ \citep{kellermann89}.

The VLASS images also enabled us to explore the radio structure of the
selected sources. We defined as extended sources i) those for which
the catalog of VLASS components\footnote{Available at {\sl
    https:\slash\slash cirada.ca\slash vlasscatalogueql0}} returned a
measurement of their major axis, $r_{\rm maj.}$, that was larger than twice the
median value of the components in the same range of declination, that
is, 2\farcs93, or ii) those in which the visual inspection of the VLASS
images showed multiple components. Based on this
criterion, three HzRQs are extended and the remaining 26 are
compact. We also defined the largest angular size of the extended
sources as the maximum separation between the farthest radio
components.  In Fig. \ref{vlass} we present the radio images of the
three extended sources: Two of them have a clear double structure. The
last source, SDSS~034402.85-065300.6 (see the left panel in
Fig. \ref{vlass}), might be interpreted as a core-jet source, but the
off-nuclear source might also be unrelated to the QSO, although there
is no optical counterpart at this location.

We explored the properties of the compact sources in more detail and show in Fig. \ref{size} the distribution of $r_{\rm
  maj.}$. Most of them are apparently unresolved sources, with $r_{\rm
  maj.}<3\farcs3$, but a tail of marginally resolved sources is also
present. The distribution of deconvolved sizes has a median of
1\farcs1, corresponding to $\sim 8$ kpc at the median redshift of the
sample.

\begin{figure*}
    \includegraphics[width=0.32\textwidth]{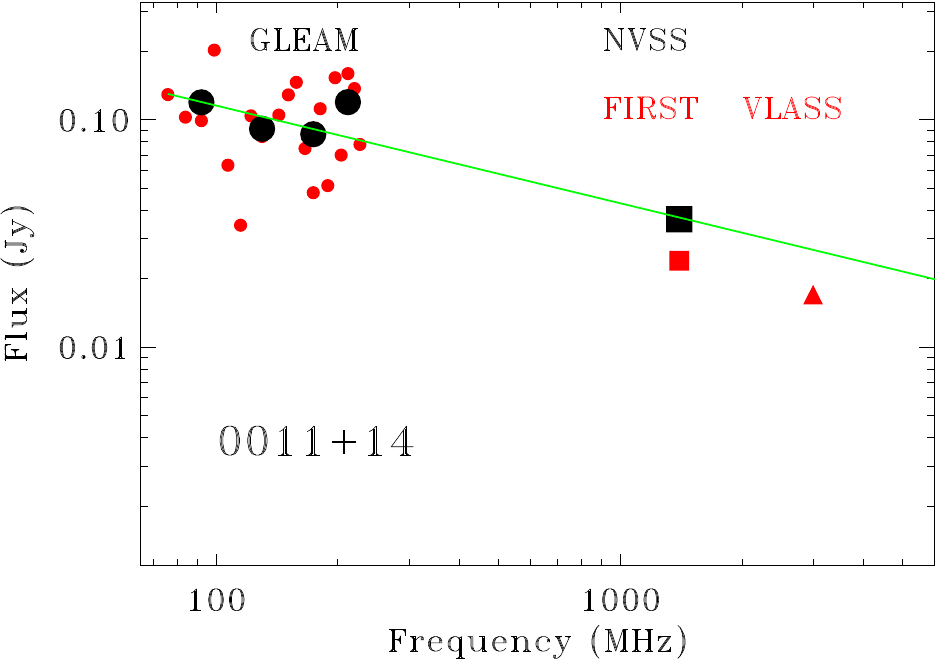}
    \includegraphics[width=0.32\textwidth]{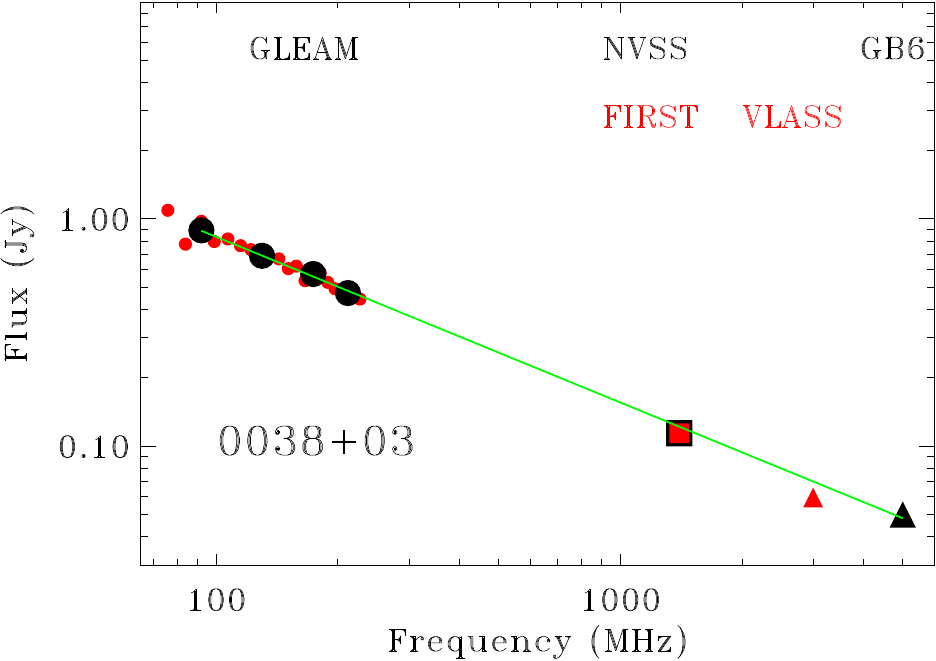}
    \includegraphics[width=0.32\textwidth]{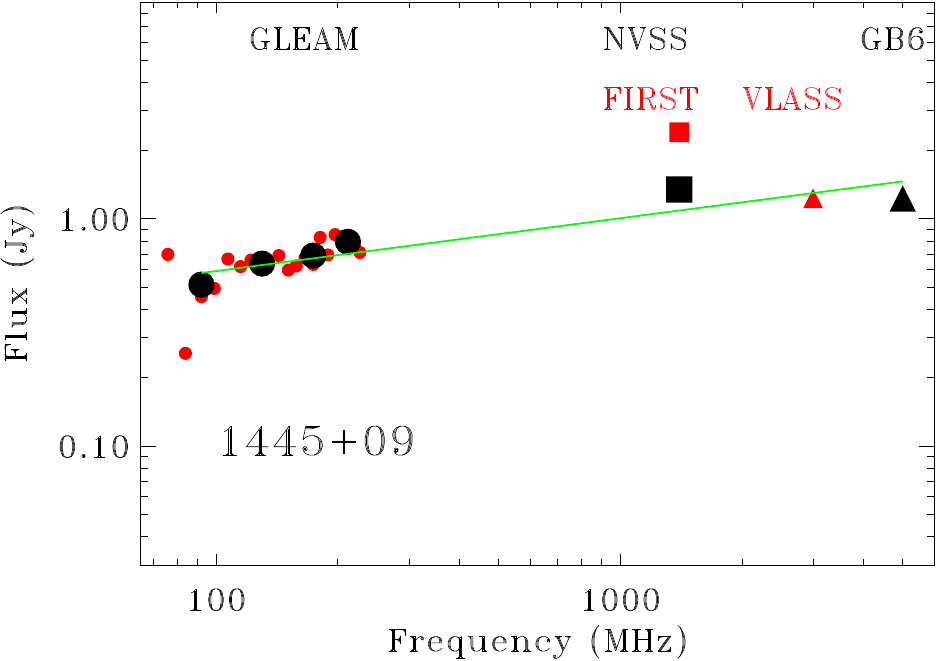}
    \caption{Three examples of radio spectra. The low-frequency regime
      is covered by the GLEAM data (small red circles). Because the errors are relatively large, we also provide flux densities by
      rebinning five frequency channels (large black circles). The other
      flux densities are taken from the NVSS (black squares), FIRST (red
      squares), VLASS (red triangles), and GB6 (black triangles). Only
      data with a similar spatial resolution (i.e., from GLEAM, NVSS, and
      GB6, all marked with black symbols) are used to obtain the
      linear fit, which is represented by the green line.}
    \label{spectra}
\end{figure*}

\section{The radio spectra}

For the 29 selected QSOs, we gathered radio data at different
frequencies. The GLEAM EGC provided us with 20 measurements from 76 to
227 MHz, which we complemented with the flux densities at 1.4 GHz from
the Faint Images of the Radio Sky at Twenty centimeters survey,
(FIRST; \citealt{becker95,helfand15}) and the National Radio Astronomy
Observatory Very Large Array Sky Survey (NVSS; \citealt{condon98}), at
3 GHz from the VLASS, and the 5 GHz from the Green Bank 6 cm survey
(GB6; \citealt{gregory96}). The cross-match for the NVSS and GB6
  was performed by adopting a search radius of 15\arcsec\ from the
  location of the VLASS source, while a search radius of 3\arcsec\ was
  used for FIRST.

Three examples of the resulting radio spectra are presented in
Fig. \ref{spectra}, and all radio spectra are presented in the
appendix.

\begin{figure}
    \includegraphics[width=0.48\textwidth]{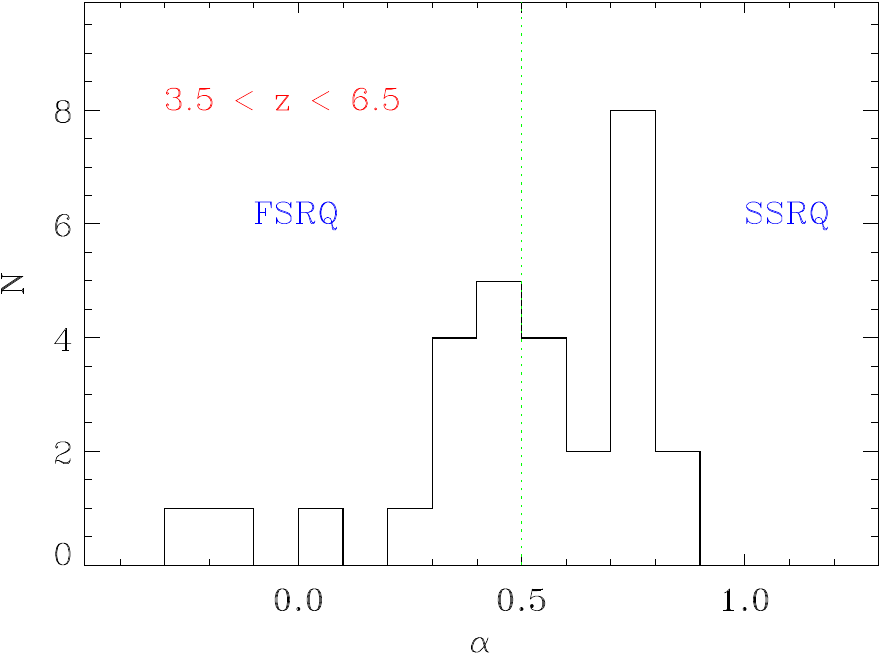}
    \caption{Distribution of spectral indices for the QSO with $3.5 <
      z < 6.5$. The vertical dashed green line separates flat- from
      steep-spectrum sources, that is, FSRQs from SSRQs.}
    \label{alfa}
\end{figure}

The radio spectra are generally well represented by a single power
law in the form $F_{\nu} \propto \nu^{-\alpha}$ over the whole frequency range. We derived the radio spectrum index $\alpha$ with a
weighted linear fit in logarithmic units. For the fit, we considered only
the radio flux density measurements obtained at similar spatial
resolution (i.e., from GLEAM, NVSS, and GB6). The high-resolution
  data from FIRST and VLASS are shown in the radio spectra to verify
  their consistency with the low-resolution data, but were excluded
  because these surveys might be missing extended structures. The
spectral indices are listed in Table \ref{tab1}, and their distribution
is shown in Fig. \ref{alfa}. The highest measured value of the spectral
index is $\alpha=0.87$, and therefore, we did not find any ultrasteep radio
sources associated with HzRQs, which agrees with the results of
\citet{Sotnikova21}.

By using $\alpha=0.5$ as a threshold to separate flat from steep
spectra, we find following the definition of \citet{condon84}
that 15 QSOs (45\%) have flat radio spectra. All of them are unresolved
sources at the VLASS spatial resolution. Several of these FSRQs show a
relatively large scatter from the power-law fit; for example,
SDSS~144516.46+095836.0 shown in the right panel of
Fig. \ref{spectra}. This is likely to be due to variability in their
radio emission, as expected from objects that are amplified by Doppler
boosting. None of the FSRQs shows clear evidence of a
steepening radio spectrum at low frequencies, as would be
expected in case of a significant contribution from diffuse, steep
spectrum emission. There is a marginal increase in the fraction of
flat sources, $f_{\rm FSRQ}$, within this redshift range: The redshift of
the eight FSRQs among the 14 objects (57\%) is higher than the
median value. However, a test of equal proportions
  \citep{wilson27} indicates that this difference is not statistically
  significant.

From the fit to the radio spectra, we estimated the luminosity of the
29 QSOs at a fixed rest-frame frequency of 500 MHz, $P_{500}$. We find
that $P_{500}$ ranges from $\sim 0.7 \times 10^{28} \WHz$ to $\sim 40
\times 10^{28} \WHz$, with a median of $\sim 4.6 \times 10^{28}
\WHz$. 

\section{Biases in the sample selection}
The selection of the sample at low frequency introduces a bias as it
might miss sources with a turnover at low frequency: This is the case
of the compact steep-spectrum sources (CSS), which are thought to be young radio
sources, in which the low-frequency emission is depressed due to self-absorption, for example (see \citealt{odea21} for a review of the
properties of CSS).

\begin{figure}
    \includegraphics[width=0.48\textwidth]{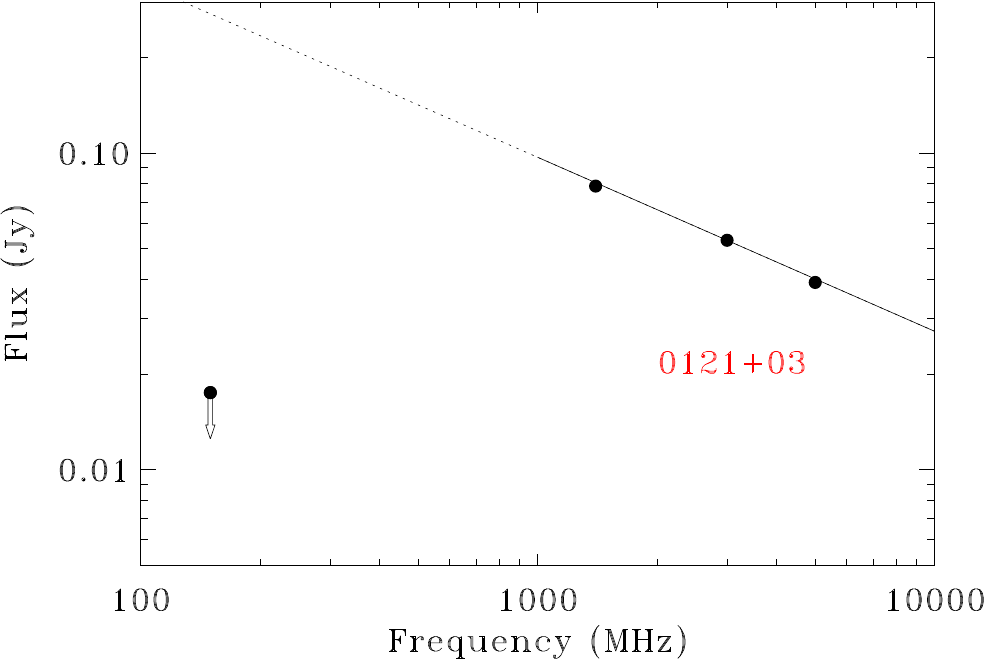}
    \caption{Example of a source with a turnover at low frequency, a
      likely CSS, missed by our selection at low frequency. The solid
      line is the best line fit to the high-frequency data. The
      extrapolation of the fit at low frequency (dotted line) is higher by a
      factor $>$10 than the upper limit from the TGSS.}
    \label{css}
\end{figure}

We investigated this issue in more detail and included the quasars
with $z>3.5$ and $\delta < 30^\circ$ in the EGC area.
We selected the QSOs with a detection in FIRST. CSS at high
  redshift are expected to be unresolved in the FIRST images (the
  spatial resolution of FIRST at z$ \sim 3.5$ is $\sim 35$ kpc), and we
  do not expect that they miss radio flux from extended
  structures. We estimated their radio power at the rest-frame
frequency of 5 GHz, $P_{5000}$, (above the typical turnover frequency
of CSS) and considered those with $L_{5000} > 4 \times 10^{27} \WHz$,
the minimum luminosity at this frequency of the selected sample of 29
quasars extracted from the EGC in the previous section. We then
derived their radio spectra by including the flux densities from TGSS,
FIRST, VLASS, and GB6. We find that a large fraction of them (38 out
of 54) has a flat spectrum, which is inconsistent with a CSS
identification. Other sources have a steep spectrum and are detected
by the TGSS, but their flux density is lower than the EGC threshold: These are
correctly not included in our sample. The remaining 8 sources are
likely CSS, with a steep spectrum between FIRST and VLASS and a nondetection (or a turnover) based on the TGSS. The CSS fraction derived
from this analysis ($\sim $ 20\%) agrees with the results of
\citet{coppejans17}. In Fig. \ref{css} we show the radio spectrum of
one such source, in which the flux density upper limit from the TGSS
lies below the measured value at 1.4 GHz. The inclusion of these
(intrinsically) steep sources in the sample might reduce $f_{\rm
  FSRQ}$ to 35\%.

\begin{figure*}
    \includegraphics[width=0.98\textwidth]{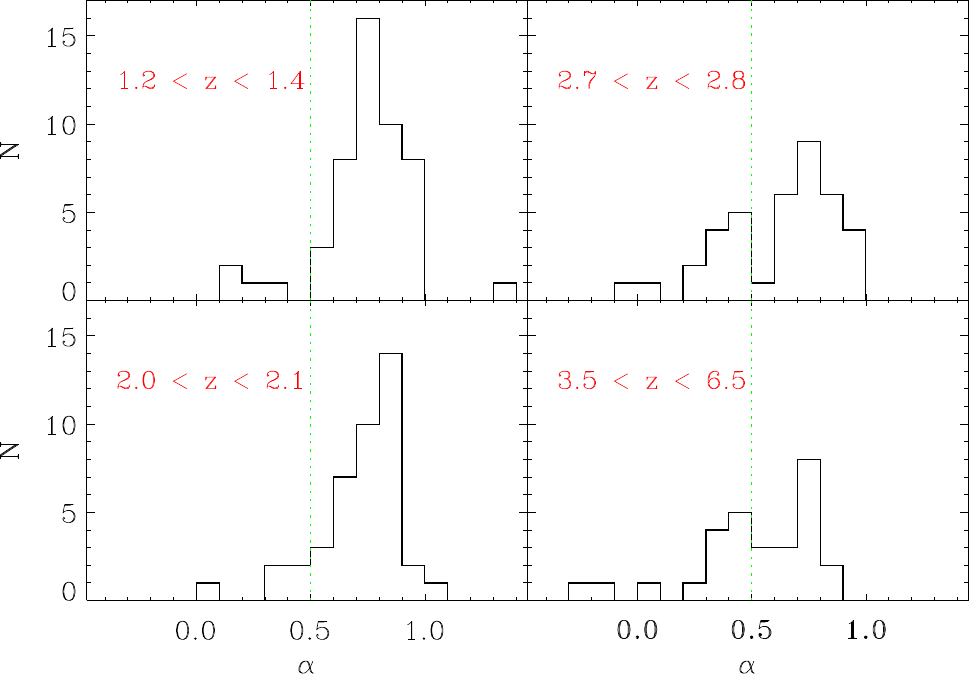}
    \caption{Distribution of spectral indices for powerful radio
      quasars ($L_{500} > 10^{28} \WHz$) in the four redshift bins
      considered (the bottom panel reproduces the same
        distribution as in Fig. \ref{alfa} for the $z>3.5$ QSOs).
      The vertical dashed green line separates flat- from steep-      spectrum sources. The fraction of FSRQS increases at
      increasing redshift.}
    \label{alfa2}
\end{figure*}

\begin{figure}
    \includegraphics[width=0.48\textwidth]{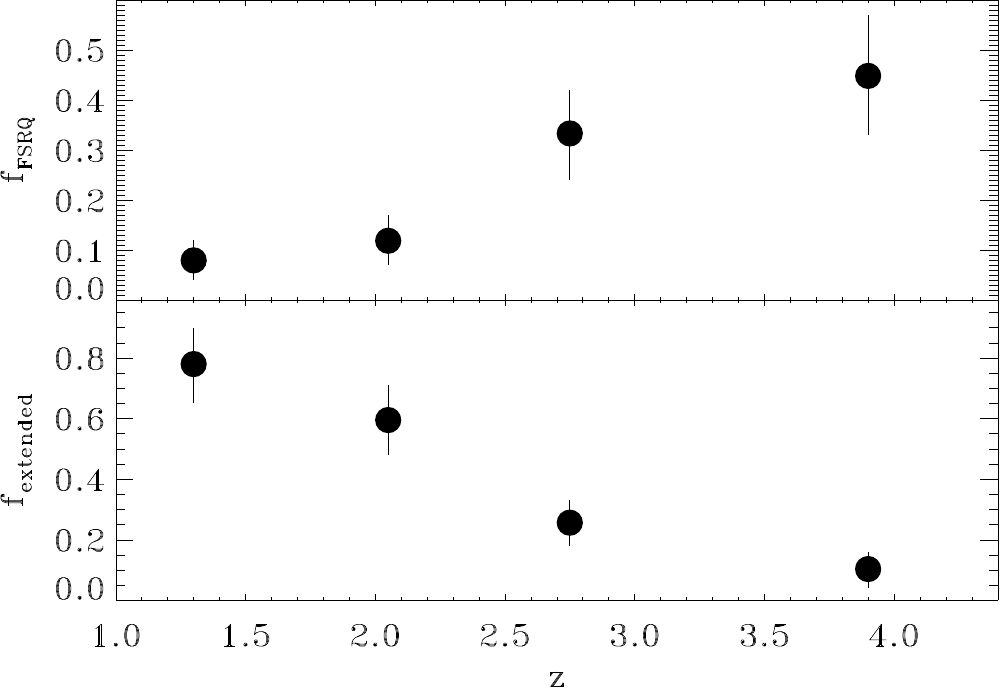}
    \caption{Evolution of the fraction of FSRQ, $f_{\rm FSRQ}$ (top),
      and of the fraction of extended radio sources (bottom) with
      redshift. The uncertainties are estimated by adopting the
      Poisson statistic.}
    \label{summary}
\end{figure}

\begin{figure}
    \includegraphics[width=0.48\textwidth]{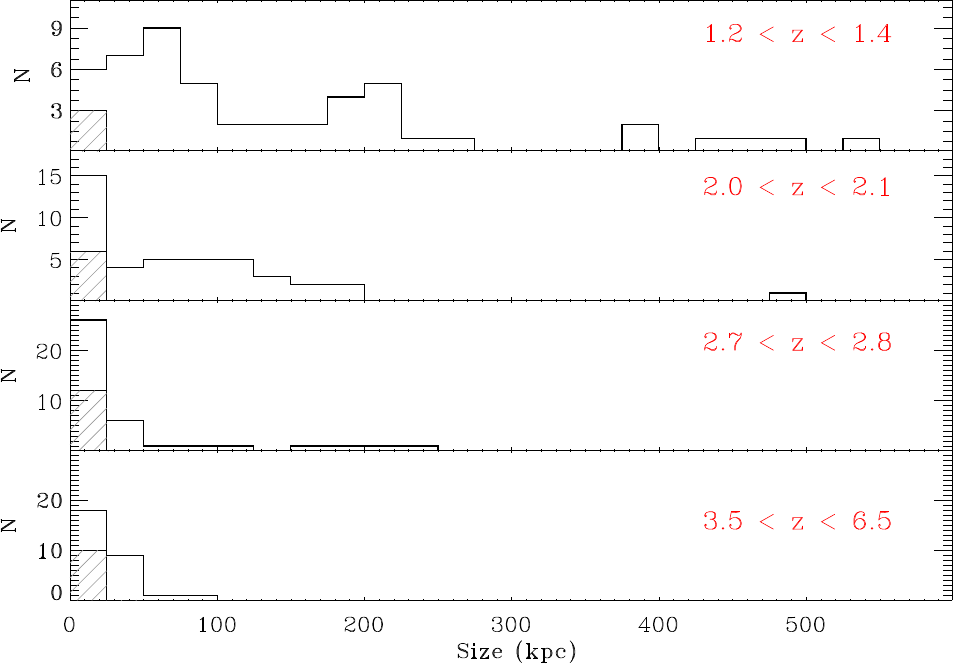}
    \caption{Evolution of the distribution of the deconvolved
      sizes of the radio sources, $f_{\rm extended}$ with
      redshift. The dashed area of the histograms represents the
        upper limits to the size of the unresolved sources.}
    \label{sizes}
\end{figure}

\section{Selection and analysis of the lower-redshift samples}

In order to set our results in a broader context, we selected three
further samples of quasars at lower redshift, namely at $1.2 < z <
1.4$, $2.0 < z < 2.1$, and $2.7 < z < 2.8$.  These three redshift bins
were selected to uniformly cover the range from the well-studied radio source samples (up to z$\sim$1) to the high-redshift sources
(z$>$3.5). The bin sizes were chosen to include a sufficient
number of sources for a robust statistical analysis.  The selection
method was the same as we used for the highest-redshift QSO, that is,
a cross-match between SDSS QSO and GLEAM sources followed by the
verification of the radio-optical association with the VLASS
images. However, for these samples, we only chose QSOs with a
radio power $P_{500} > 10^{28} \WHz$ for consistency with the
luminosity range of the HzRQ sample. The total number of sources in
the various redshift bins is 50 for $1.2 < z < 1.4$, 42 for $2.0 < z <
2.1$, and 39 for $2.7 < z < 2.8$.

We estimated their radio spectral indices. The distributions at the
different redshifts are compared in Fig. \ref{alfa2}. There is a
  marked difference in these distributions. Sources with a higher
  redshift have flatter spectra in general. More quantitatively, in the top panel of
Fig. \ref{summary}, we show the evolution of the fraction
of FSRQs (i.e., sources with $\alpha < 0.5$). It increases with
redshift from $\sim 8\%$ at $z\sim 1.3$ to $\sim 45\%$ at $z> 3.5$
(see Table \ref{tab2} for a summary of the results).

Following the same strategy as adopted in Sect. 3.1, we searched for
  candidate CSS in these lower-redshift samples. We found 4, 5, and 4
  such sources at $1.2 < z < 1.4$, $2.0 < z < 2.1$, and $2.7 < z <
  2.8$, respectively. The corresponding CSS fractions are similar
  (ranging from 7\% to 10\%) and smaller than we found for the highest
-redshift sources ($\sim 20\%$). However, the small number statistics
  prevents us from drawing any strong conclusion about the evolution of the CSS
  fraction with redshift.

  The fraction of extended sources instead decreases with redshift,
  from $\sim 80\%$ to $\sim 15\%$. This latter effect is not due
    to the different redshift ranges of the sources considered
    because with the adopted cosmological parameters, the scale varies
    from 7.4 kpc arcsec$^{-1}$ at z=3.5 to 8.5 kpc arcsec$^{-1}$ at
    z=1.3. In both cases, a sharp transition appears to
  occur at $z\sim 2.5$.  The sizes of the radio sources also show
  a strong evolution with redshift (see Fig. \ref{sizes}): While in the
  lowest redshift bin we considered, the size of about half of the QSOs
  exceeds $\sim 100$ kpc, no QSO with $z>3.5$ is associated with
  such a large radio source.

\begin{table*}[h]
  \caption{Results obtained for the various QSO samples
    considered. } \centering
    \begin{tabular}{c | c r r | c r }
      \hline
& \multicolumn{3}{c}{Bright sample } & \multicolumn{2}{c}{Faint sample}\\
& \multicolumn{3}{c}{$P_{500} \gtrsim 10^{28} \WHz$} & \multicolumn{2}{c}{$P_{500} \gtrsim 10^{27} \WHz$}\\
      \hline
      z & N & Ext. & FSRQ & N & FSRQ \\
      \hline
 $1.2 < z < 1.4$ & 50 & 39 &  4 (8$\pm4$\%)  & 241 & 36 (15$\pm3$\%)  \\  
 $2.0 < z < 2.1$ & 42 & 25 &  5 (12$\pm5$\%)       & 145 & 40 (28$\pm4$\%)  \\  
 $2.7 < z < 2.8$ & 39 & 10 & 13 (33$\pm9$\%)       & 177 & 65 (37$\pm5$\%)  \\  
 $3.5 < z < 6.5$ & 29 &  3 & 13 (45$\pm$12 \%)       & 172 & 81 (47$\pm5$\%)  \\  
  \hline
    \end{tabular}
    \label{tab2}
    \medskip
    
    \noindent
{\small Summary of the results obtained for the bright sample (left
  columns; see Sect. 3) and the lower-luminosity faint sample (right
  columns; see Sect. 5). We report the redshift range, the number of
  sources, the number (and fraction) of sources with a flat
  ($\alpha<0.5$) radio spectrum, and for the bright sample alone, the
  number of extended radio sources. The uncertainties are estimated by
  adopting the Poisson statistic.}
\end{table*}

\section{Exploring the properties of quasars with a lower radio luminosity}

The selection of sources detected in the GLEAM observations sets a
lower limit to the radio luminosity of the quasars considered of
$P_{500} \gtrsim10^{28} \WHz$. These sources, to which we refer as the bright sample, then represent the extreme high-power end
of the population of radio quasars. We here explore the properties of
quasars with a lower radio luminosity to test whether the results obtained
above apply more generally to the overall population and to set them
on a stronger statistical basis.

The selection was again performed at low frequency, but using the
images produced by data release 2 of LOFAR LoTSS, which covers
$\sim$ 5,700 square degrees.  The advantage of using LoTSS data is that their depth
exceeds that of the GLEAM images (the LoTSS
reaches a $\sim$80\% completeness at 0.8 mJy), at the expense of losing
the information on the full low-frequency radio spectrum. 
  However, having established that a power law provides a good
  representation of the radio spectra of HzRQs, we can rely on the
  LOFAR data, combined with higher-frequency measurements, to separate
  SSRQs from FSRQs.

\begin{figure}
    \includegraphics[width=0.48\textwidth]{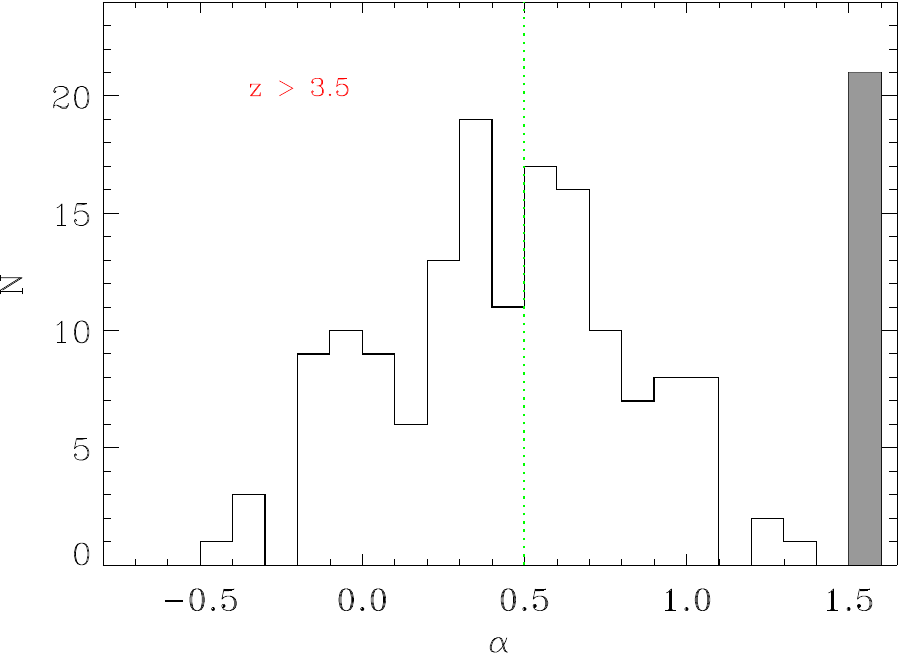}
    \caption{Distribution of the spectral indices obtained with a linear
      fit to the LOFAR, FIRST, and VLASS data to the sources of the
      faint sample, that is, with $P_{500} \gtrsim 10^{27} \WHz$. The
      gray bar represents the 21 sources that were not detected by in the high
    -frequency surveys. The vertical dashed green line separates
      FSRQs from SSRQs.}
    \label{lofar}
\end{figure}

We selected the quasars with $z>3.5$ that were covered by both LoTSS and
FIRST. We adopted a threshold for the flux density at 150 MHz of 5
mJy. With this restriction, we can set a lower limit on their spectral
index between 150 MHz and 1.4 GHz for all quasars that were not detected by
FIRST $\alpha_{150,1400}\gtrsim 0.7$, and they can be safely classified
as SSRQ. We selected 172 quasars, to which we refer to as
  the faint sample, 21 of which do not have a FIRST counterpart to
the LOFAR source. At this stage, we also included the VLASS flux
  density measurements. By assuming that their radio spectra are
reproduced with a single power law, as seen in the high-power sources,
we estimated the spectral slopes with a weighted linear fit in log-log
space including measurements from LOFAR, FIRST, and VLASS. The
distribution of spectral indices we obtained is shown in
Fig. \ref{lofar}. The fraction of FSRQ obtained with this method is
47\%, which is similar to that found for the bright sample.

Following the same strategy as adopted in the previous section, we
also explored the behavior of quasars at lower redshift by estimating
their radio spectral slopes. We limited the analysis to sources within
the same luminosity range as for the higher redshift bin, that is, $
P_{500} > 10^{27} \WHz$. The results are summarized in
Table \ref{tab2}. The evolution of $f_{\rm FSRQ}$ with redshift is
similar to that observed in the bright sample, although the
  fraction of FSRQs appears to decrease less sharply with redshift.

\section{Discussion}

\subsection{Evolution of radio spectral indices and the fraction of obscured RLAGN}

The main result of our analysis is the strong increase in the fraction
of FSRQ, $f_{\rm FSRQ}$, with redshift. FSRQs are radio sources in
which the highly relativistic jets form a small angle with the line of
sight, and their emission is amplified by the effects of Doppler
boosting.  The fraction of FSRQs within a population of radio sources
can be estimated as $f_{\rm FSRQ} \sim 1/(2\Gamma^2)$, where $\Gamma$
is the jet bulk Lorentz factor. We consider as highly boosted sources
all those seen at an angle $\theta \lesssim 1/\Gamma$. This is a
reasonable assumption considering the strong dependence of the Doppler
boosting factor on the viewing angle. By reversing this argument, it
is possible to estimate the number of sources expected to be the
misaligned counterparts (i.e., HzRGs and SSRQs) of a given sample
FSRQ.

However, for a flux-limited sample, or for sources
that were selected based on the observed radio luminosity, this line of
reasoning does not hold, as already pointed out by
\citet{lister19}. This is because FSRQs might exceed the threshold of
the sample only because of the contribution of the nuclear beamed
emission. Their isotropic radio emission, which is mainly produced by the
extended structures, might instead be significantly lower than those
of the SSRQs selected at the same flux density (or luminosity)
threshold.  Considering the steepness of the luminosity function of
radio-loud AGN, a flux-limited sample includes a large number of FSRQs
with a lower isotropic luminosity that that of the SSRQs.

We then followed a different approach and considered samples of radio
quasars that covered the same range of radio luminosity at different
redshifts. When we assume that the beaming factor and the slope of the
luminosity function do not change with redshift, the fraction of FSRQs
within the population of radio sources (including both narrow- and
broad-line object) above a given observed luminosity are not expected to
vary. The observed increase in $f_{\rm FSRQ}$ with redshift within the
quasar population is due to a corresponding decrease in the
  fraction of SSRQs.

For the most powerful quasars, we find $f_{\rm FSRQ} \sim 8\%$ for $z
\sim 1.2$ and $f_{\rm FSRQ} \sim 45\%$ for $z>3.5$. When we consider the
likely CSS sources (as discussed in Sects. 3.1 and 4), this fraction
might be reduced to $f_{\rm FSRQ} \sim 13/(29+8) \sim 35\%$ in the
high-redshift sample and to $f_{\rm FSRQ} \sim 7\%$ in the
  lowest redshift bin.

When we consider the whole population of RLAGN at low redshift, they
split almost equally into obscured and unobscured sources. The
fraction of beamed objects is significantly smaller, $\sim 1/(2\Gamma^2)$. As a result of Doppler boosting, however, this
fraction increases in a flux luminosity-limited sample to the
observed value of $\sim 8$\% of the QSOs at z$\sim 1.3$. These FSRQs
are sources of lower isotropic radio luminosity that meet the selection
due the amplification of their nuclear emission. Under the assumptions
i) that the beaming factor and the slope of the luminosity function at
high luminosities do not change with redshift and ii) that a purely orientation AGN unified model is valid, the fraction of FSRQs with
respect to the whole population of RLAGN is expected to be
constant. This is because in the scenario we assumed,  the beamed sources are drawn from the same range of isotropic luminosity and are
boosted by the same factor. At high redshift, the fraction of obscured
sources is not known because only the unobscured sources (the QSOs) are
visible. However, the observed increase in $f_{\rm FSRQ}$ to $\sim 45\%$
at $z>3.5$, which is a factor $\sim 5$ higher than at $z \sim 1.3$ (and the
same effect is obtained by considering the presence of CSS in both
samples), can be explained when the fraction of SSRQs decreases by the
same amount, that is, from $\sim 50\%$ to $\sim 10 \%$. The remaining
90\% of the RLAGN population are obscured in the optical band and
appear as high-redshift radio galaxies.

A similar conclusion was reached by \citet{ghisellini16}. They
suggested that at $z>4$, jetted sources are almost completely
surrounded by obscuring material in order to account for the
relatively low number of radio sources that can represent the
misaligned parent population from which FSRQs are drawn.
\citep{caccianiga24} found no evidence for obscuration in RLAGN within
angles of $\sim10^\circ-20^\circ$ from the jet axis through modeling
of their luminosity function. This is not in contrast with our
results. Several studies (see, e.g.,
\citealt{merloni14,vijarnwannaluk22}) suggested that most
high-redshift AGN are obscured. The large fraction of obscured sources
at high-z we derive from our study also extends this conclusion to the
population of radio loud AGN.

\subsection{Evolution of the size of the radio sources}

Another difference between the lower- and higher-redshift radio quasars
is the fraction of extended sources, which drops from 80\% at z=1.2 to
$\sim 15\%$ for z$>$3.5 with a strong decrease with the redshift of the
sizes of the extended sources. We tested whether this can be explained as
an orientation effect. The fraction of SSRQs statistically corresponds to a maximum viewing angle with respect to our line
of sight of $\sim 25^\circ$. The foreshortening expected for this
range of orientations causes the angular size of a source to be
reduced by an average factor of $\sim 3$.

At low redshift, the maximum viewing angle instead is $\sim 50^\circ$,
corresponding to an average foreshortening by a factor 1.8, which is only 1.4
times smaller than at higher redshift. We conclude that the smaller
angular size of the radio sources that is associated with high-redshift
quasars with respect to those at lower redshift is mainly due to an
intrinsic effect.  We argue that this is due to the
same dense gas that causes the obscuration of their nuclei, which
lows the expansion of the radio source. This idea is reminiscent
of the early suggestion that compact steep-spectrum sources are those
in which the progress of their jets is hampered by the
large amount of gas in their nuclear regions (e.g.,
\citealt{vanbreugel84}).

This interpretation suggests that the CSS fraction should
  increase with redshift. This is indeed observed, as this fraction
  increases from 7\%-10\% for z$\sim1.3-2.8$ and reaches $\sim 20\%$ at
  the highest redshift probed by our analysis.  However, the small
  number statistics prevents us from drawing any strong conclusion about the
  evolution of the CSS fraction with redshift.

\subsection{Are there ultrasteep sources among the HzRQs?}

Another interesting result is that we found no USS among the highest-redshift powerful
radio quasars, that is, sources with a radio spectrum with
a slope $\alpha > 1$, even though the selection of low
frequencies should favor the inclusion of sources like this. Conversely,
the faint sample contains 12 quasars with $\alpha > 1$ out of a total
of 91 SSRQ, which corresponds to 13\%. This fraction can be higher (up to
36\%) due to the large number of sources (21) for which the nondetection
in FIRST and VLASS does not enable us to estimate their spectral slope.

We stress this result because, as reported in the Introduction, most
HzRGs (the parent population of radio quasars) were found by obtaining
spectra of the optical counterparts of ultrasteep radio sources. The
broad-band radio spectra we obtained for the SSRQ indicate that USS
are instead quite rare even at $z>3.5$. The lower observed spectral
slopes with respect to those required for a USS classification cannot
be due to the contribution of a bright flat radio core. The signature
of this effect is a spectrum with a high $\alpha$ value at low
frequencies, followed by a sharp flattening at higher frequencies
where the core emission becomes dominant. This is not observed in any
of the high-z SSRQs.

This result implies that although this method is very efficient (the
fraction of USS that are indeed at high redshift is quite high; see,
e.g., \citealt{rottgering97}), it misses the bulk of the RLAGN
population. Furthermore, the HzRGs selected with this procedure might
be biased and may not represent the overall population.

\section{Summary and conclusions}

We explored the radio properties of a sample of 29 high-redshift
($z>3.5$) SDSS quasars selected at low frequency (76 MHz) based on
GLEAM observations. The selected sources are all very powerful radio
AGN, with a rest-frame luminosity at 500 MHz $L_{500 {\rm MHz}}\gtrsim
10^{28} \WHz$. We complemented the GLEAM data with those of other 
  radio surveys at higher frequencies, namely NVSS, FIRST, VLASS, and
  GB6, and we obtained radio spectra ranging from 76 MHz to 5 GHz. The
spectra are generally well reproduced with a single power law with a
range of slopes between $\alpha \sim -0.3$ to $\alpha \sim 0.9$
(defined as $f_\nu \propto\nu^{-\alpha}$). The sample contains
a large fraction ($\sim 45\%$) of flat-spectrum sources, that is, with
$\alpha < 0.5$.

The selection at low frequency is instrumental for building a sample
in which the radio emission is dominated by the large-scale optically
thin emission to reduce the orientation effects. However, this
choice biases the sample against the inclusion of radio sources with a
turnover at low frequency, that is, of the CSS. We estimate that nine CSS at $z>3.5$ with the same power at high frequency as the
main sample are excluded from the selection.

We studied the evolution of the QSOs by comparing the properties of
samples with a lower redshift. Again at low frequency, we selected radio
quasars with a threshold of the radio power $L_{500}\gtrsim10^{28}
\WHz$. We find that the fraction of FSRQs increases with redshift from
8\% at z=1.2 to $\sim 45\%$ at $z>3.5$, with a sharp upturn at $z
\gtrsim 2.5$. A similar result is found by studying quasar samples that were
selected from the deeper low-frequency survey LoTSS, which extended the
range of radio power down to $L_{500}\sim10^{27} \WHz$.  We interpret
this effect as due to a corresponding decrease in the fraction of
steep-spectrum quasars that becomes as low as $\sim 10\%$ at high
redshift. Under the assumptions i) that the beaming factor and the
slope of the luminosity function do not change with redshift and ii)
that a purely orientation AGN unified model is valid, this
implies that the central regions in $\sim 90\%$ of the powerful radio-loud AGN at high
redshift cannot be observed because they are
  obscured in the optical and UV bands. They can only be recognized as
quasars when their are seen within a rather small angle, $\sim
25^\circ$, from our line of sight. This supports previous suggestions
that the bulk of the high-redshift AGN population is enshrouded by a
large amount of gas and dust.

The radio structures of high-redshift quasars are also much more
often compact (with a typical size of $\sim 8$ kpc) than those of quasars of
similar power at lower redshift. We also found a trend for the size of
radio sources to decrease with increasing redshift. This change cannot simply be due to a projection effect, which only plays a marginal
role (we estimate that the average foreshortening for the high-redshift sources is only 1.4 higher than at low redshift), but it
is likely due to the very same dense gas that surrounds the more distant
sources, which slows the source expansion down.

We conclude that most high-redshift radio-loud AGN do not
manifest themselves as quasars, but as radio galaxies. This population
is still poorly studied because it is difficult to identify
these objects. As discussed in Sect. 1, the typical targets of
spectroscopic studies that searched for HzRGs are the ultrasteep radio
sources. In this study, we did not find any bright QSO with a steep
radio spectrum, and the faint sample contained only a small fraction. Under
the assumption that the radio spectra of SSRQs are representative, as
suggested by the RLAGN unifying model also of their misaligned
counterparts (i.e., of HzRGs), the general population of HzRGs currently remains largely elusive.

\begin{acknowledgements}
\end{acknowledgements}

\bibliographystyle{./aa}

\appendix
\section{Radio spectra of the 29 high-redshift quasars}

\begin{figure*}
    \includegraphics[width=0.30\textwidth]{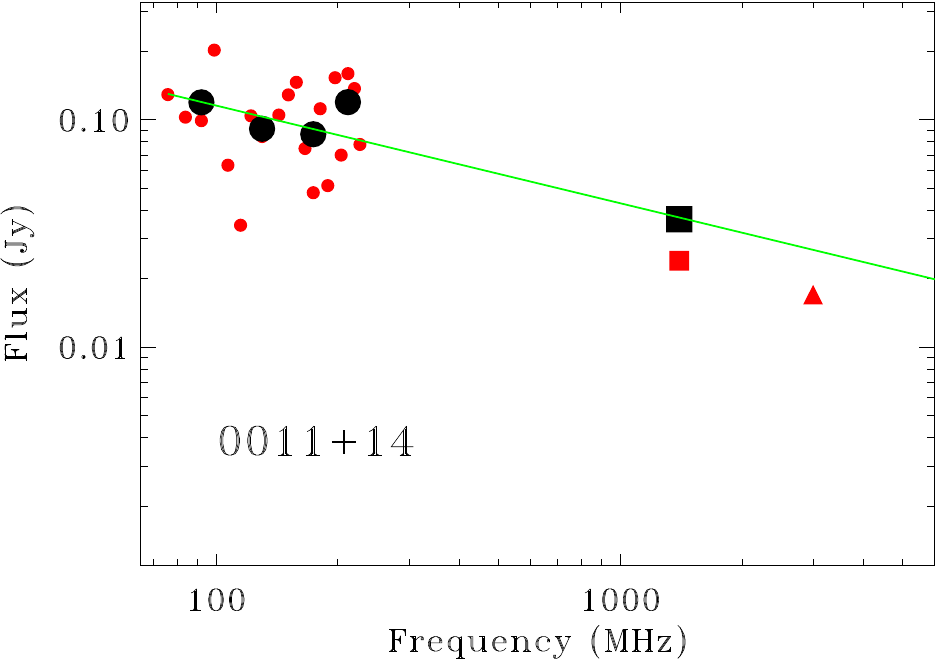}
    \includegraphics[width=0.30\textwidth]{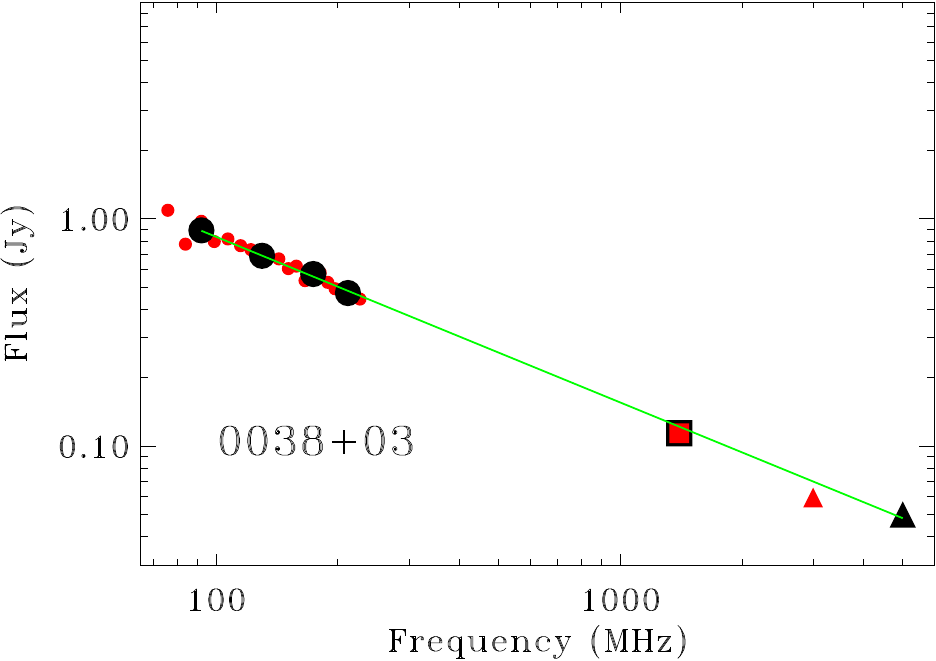}
    \includegraphics[width=0.30\textwidth]{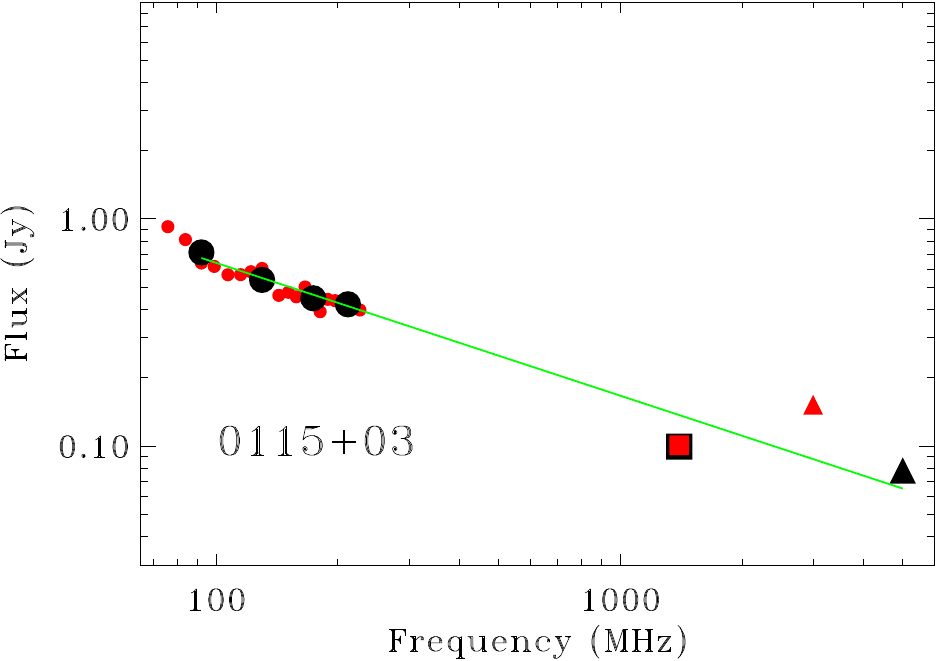}
    \includegraphics[width=0.30\textwidth]{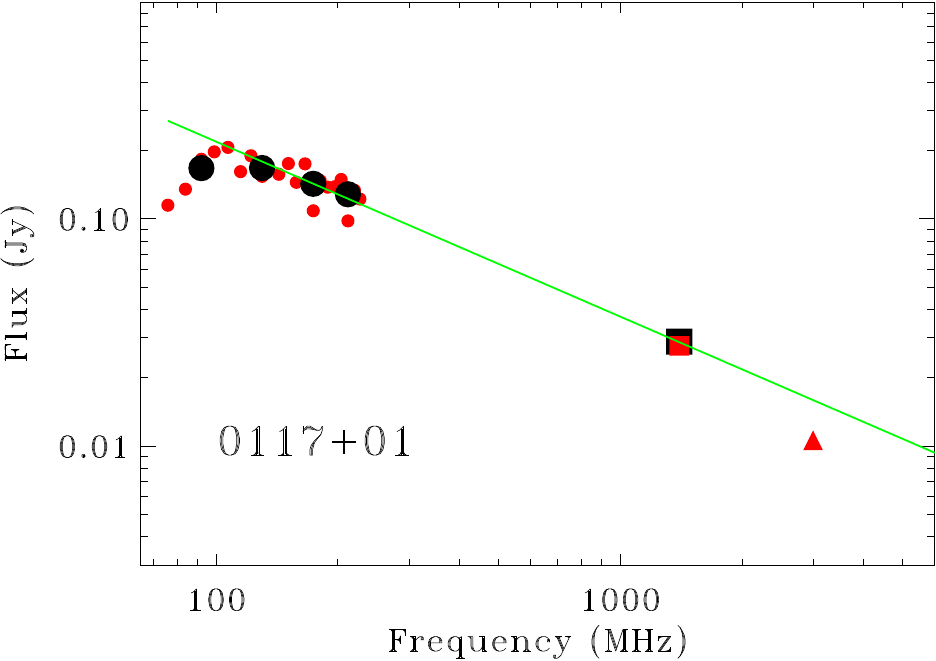}
    \includegraphics[width=0.30\textwidth]{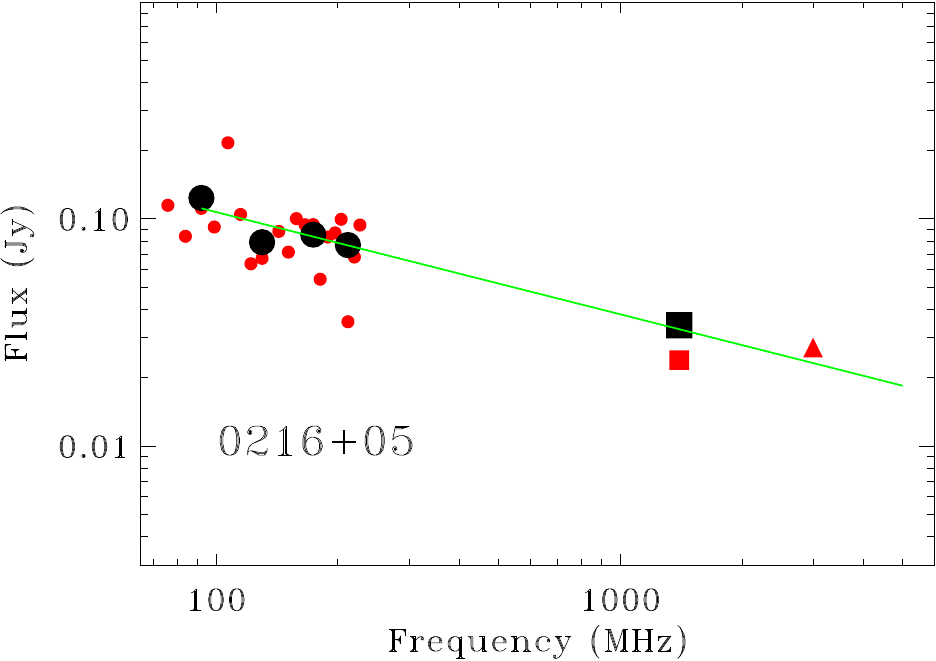}
    \includegraphics[width=0.30\textwidth]{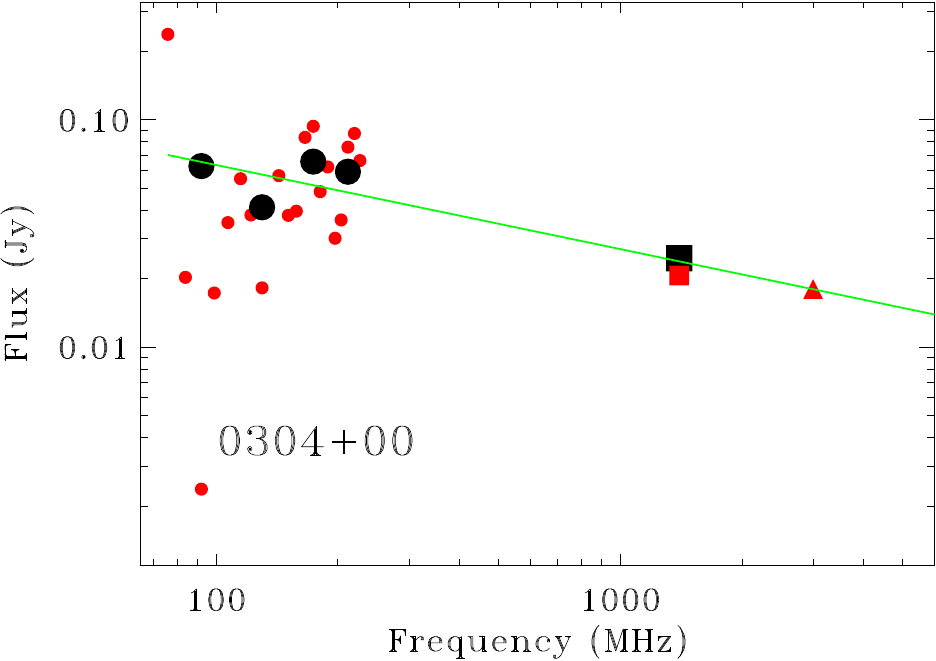}
    \includegraphics[width=0.30\textwidth]{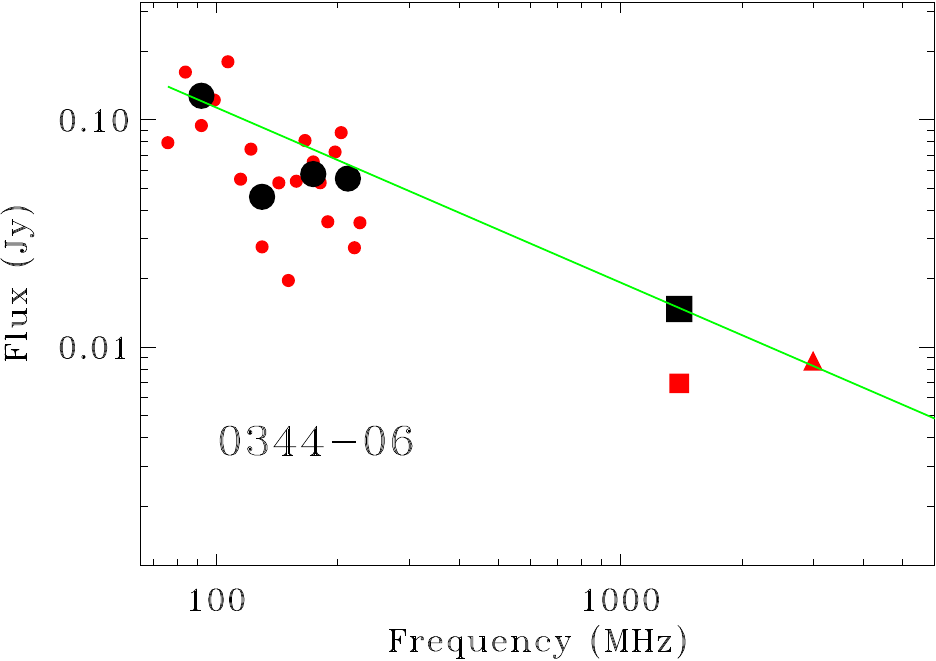}
    \includegraphics[width=0.30\textwidth]{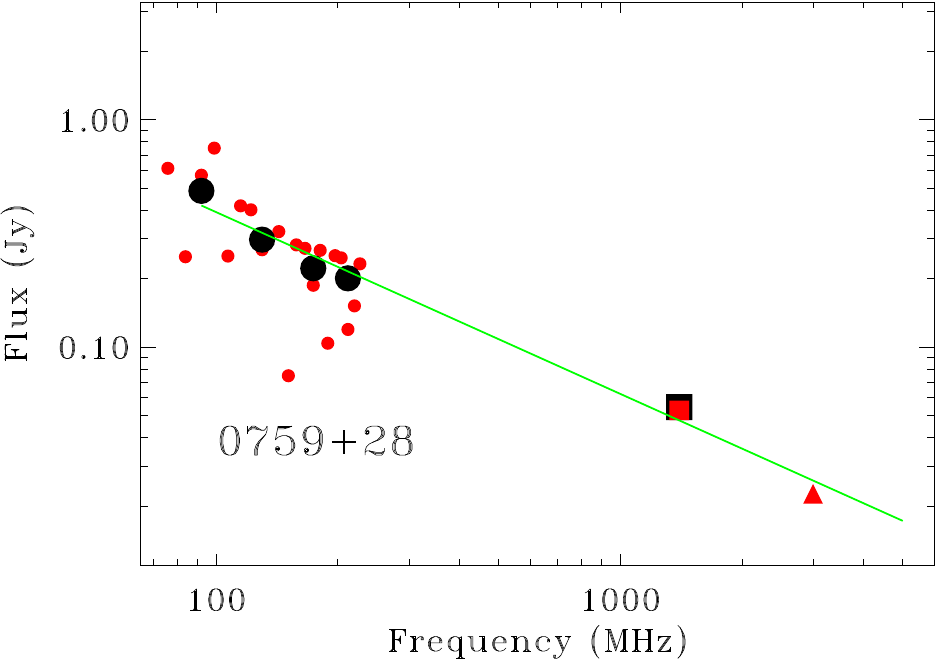}
    \includegraphics[width=0.30\textwidth]{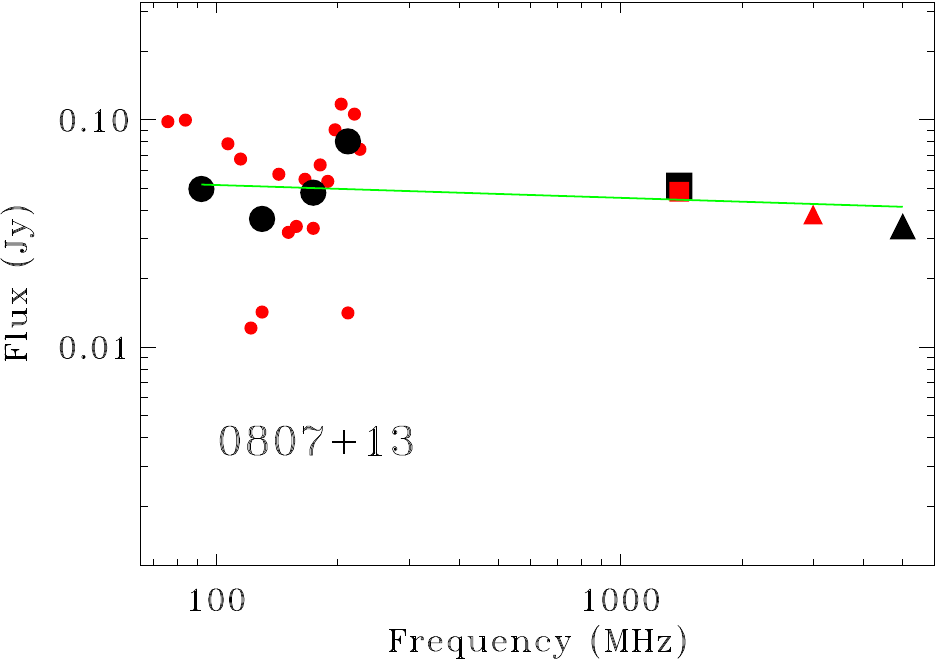}
    \includegraphics[width=0.30\textwidth]{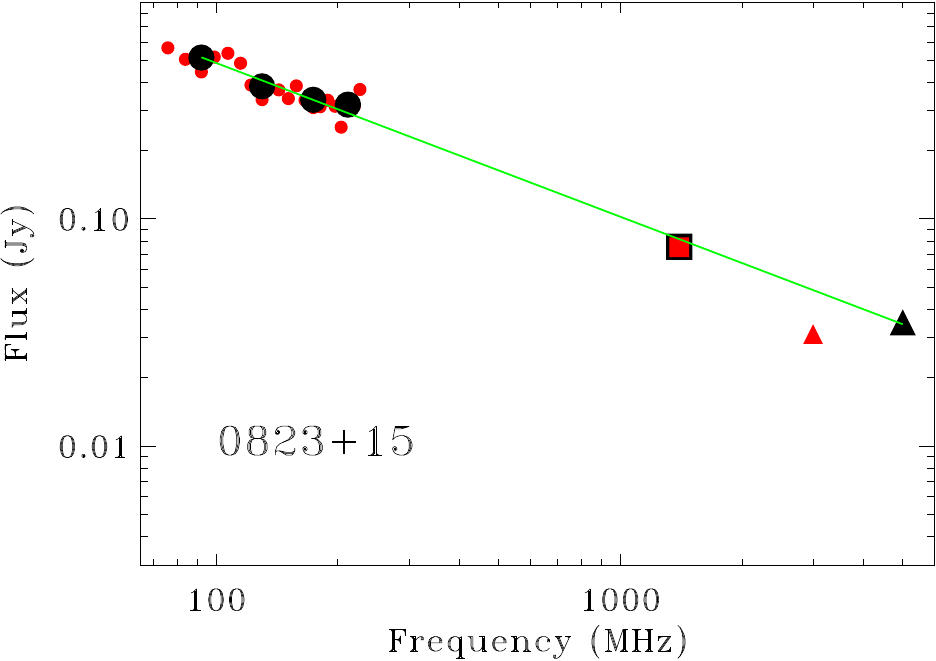}
    \includegraphics[width=0.30\textwidth]{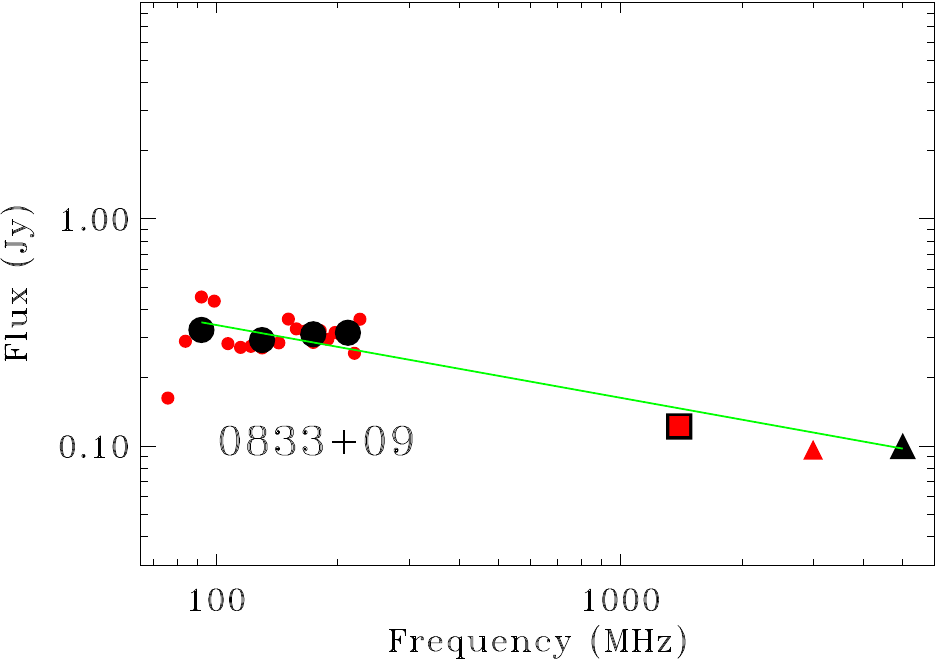}
    \includegraphics[width=0.30\textwidth]{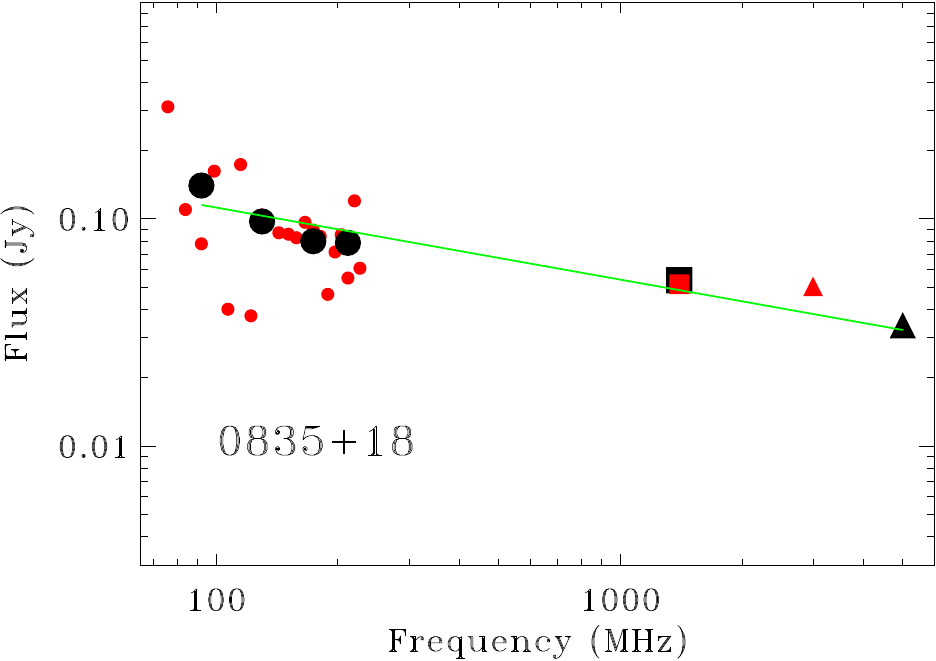}
    \includegraphics[width=0.30\textwidth]{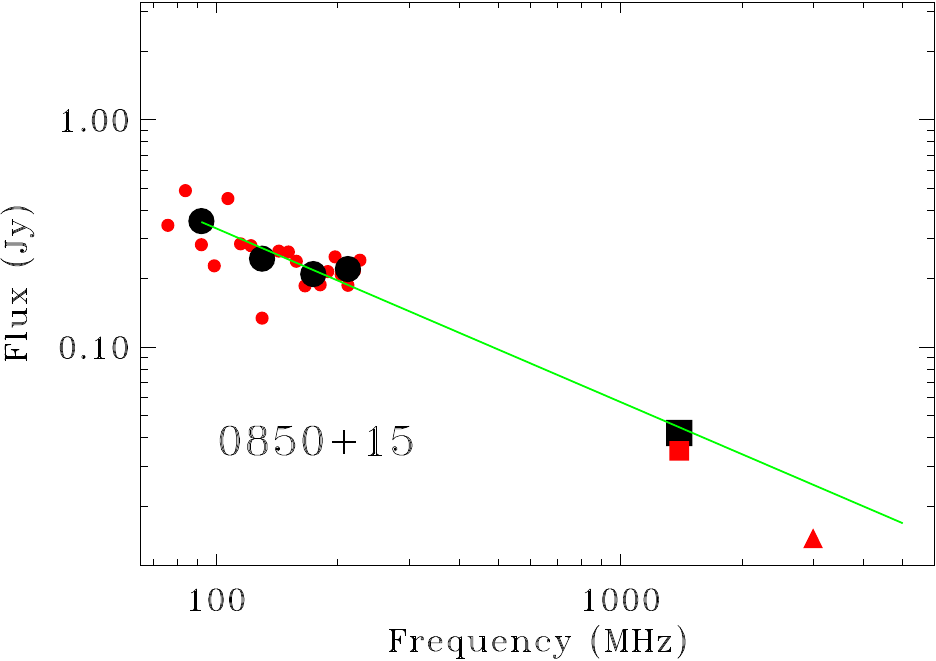}
    \includegraphics[width=0.30\textwidth]{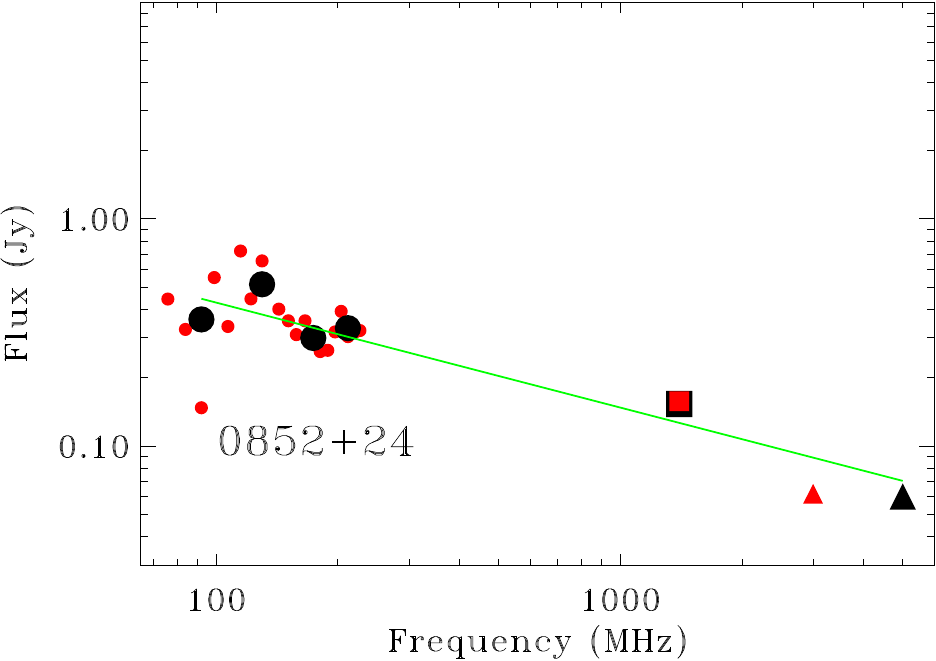}
    \includegraphics[width=0.30\textwidth]{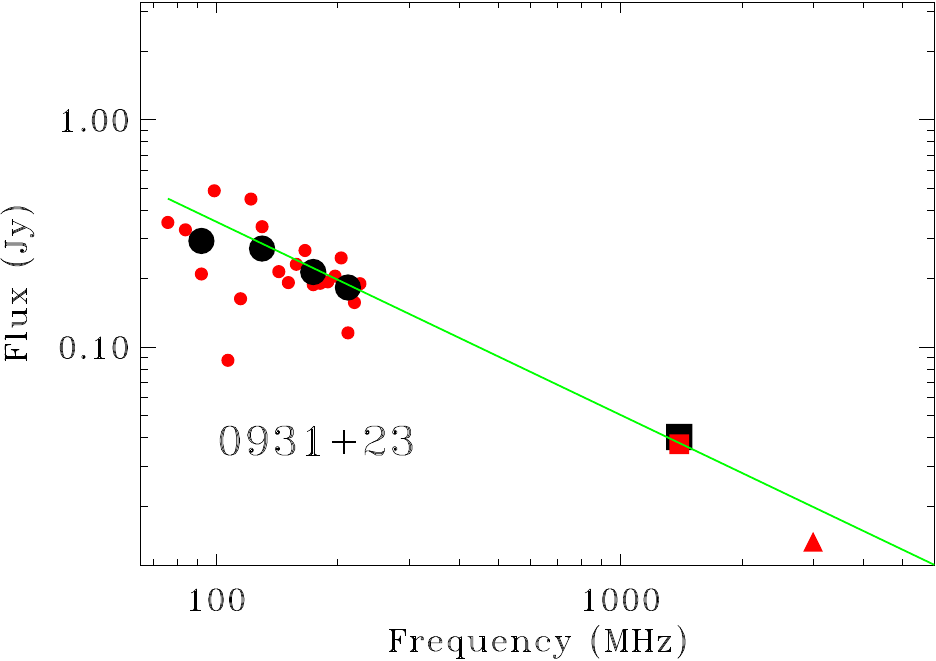}
    \includegraphics[width=0.30\textwidth]{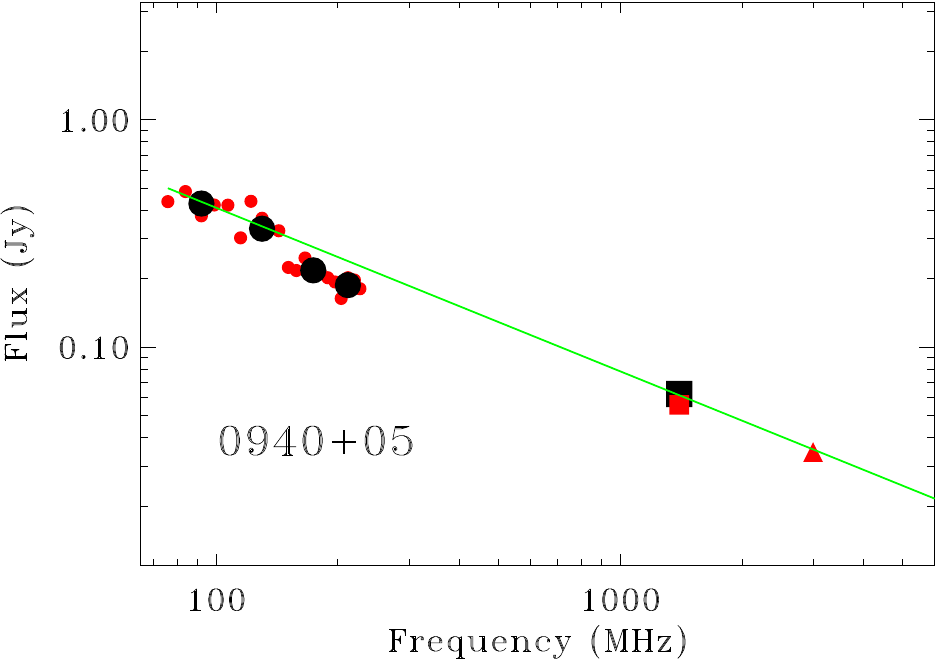}
    \includegraphics[width=0.30\textwidth]{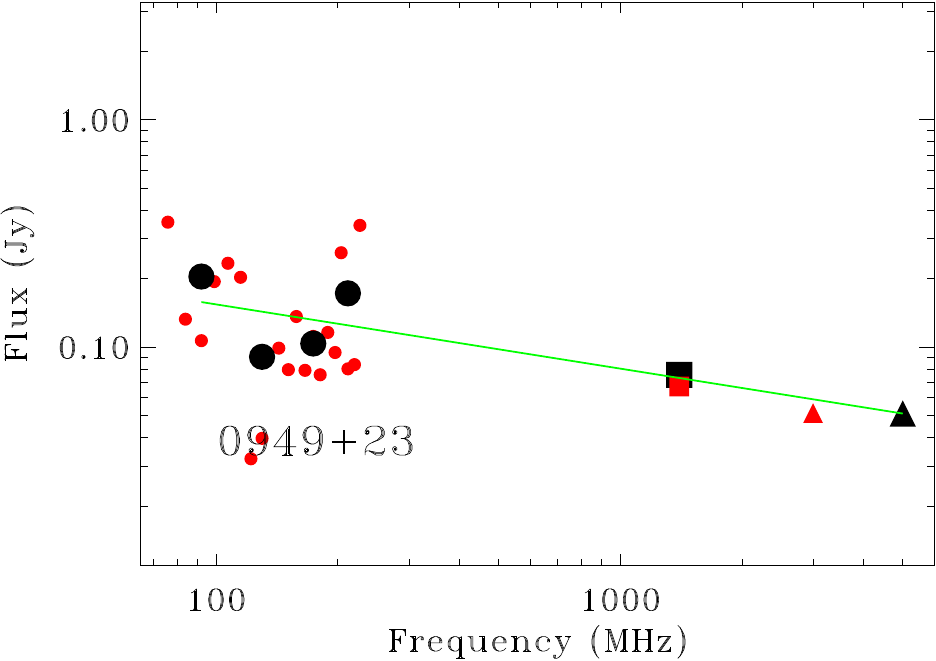}
    \includegraphics[width=0.30\textwidth]{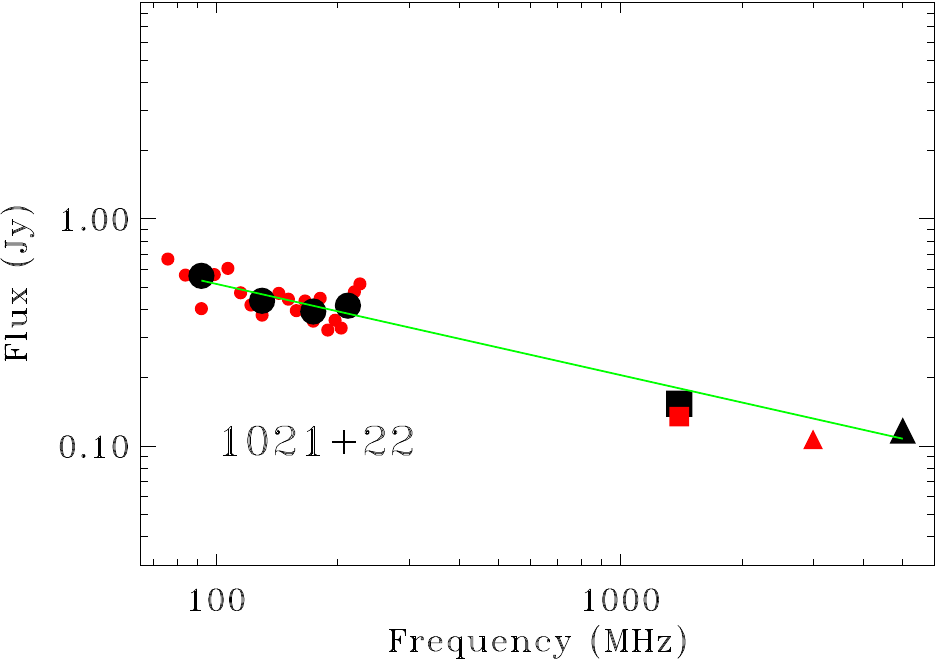}
    \caption{Radio spectra of the 29 quasars with $z>3.5$. Considering
      the relatively large errors of the GLEAM data (small red
      circles) we provide flux densities by rebinning five frequency
      channels (large black circles). The other flux densities are
      from the NVSS (black squares), FIRST (red squares), VLASS (red
      triangles), and GB6 (black triangles). }
    \label{spectra1}
\end{figure*}

\addtocounter{figure}{-1}
\begin{figure*}
    \includegraphics[width=0.30\textwidth]{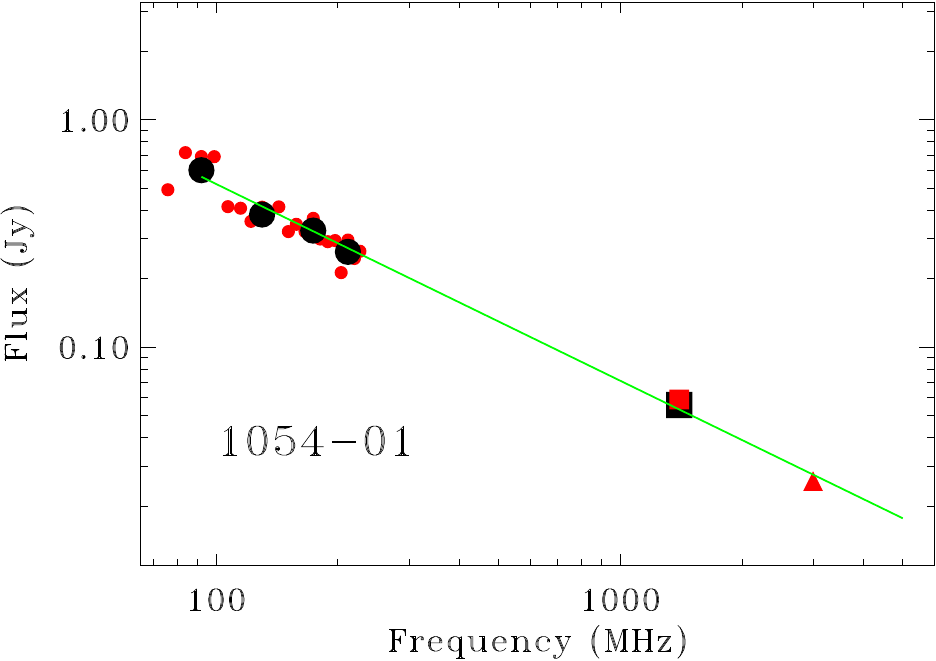}
    \includegraphics[width=0.30\textwidth]{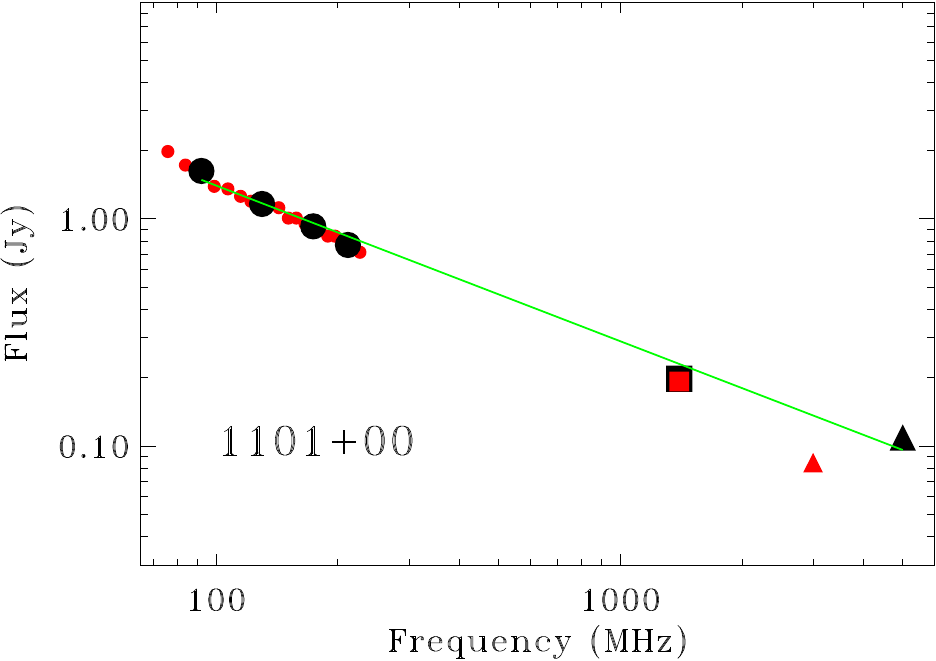}
    \includegraphics[width=0.30\textwidth]{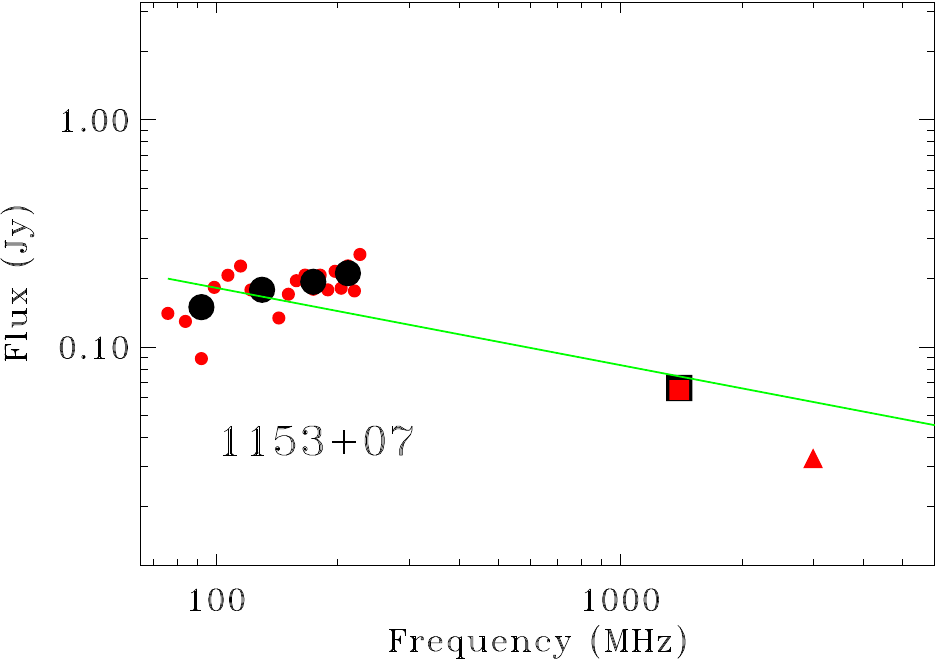}
    \includegraphics[width=0.30\textwidth]{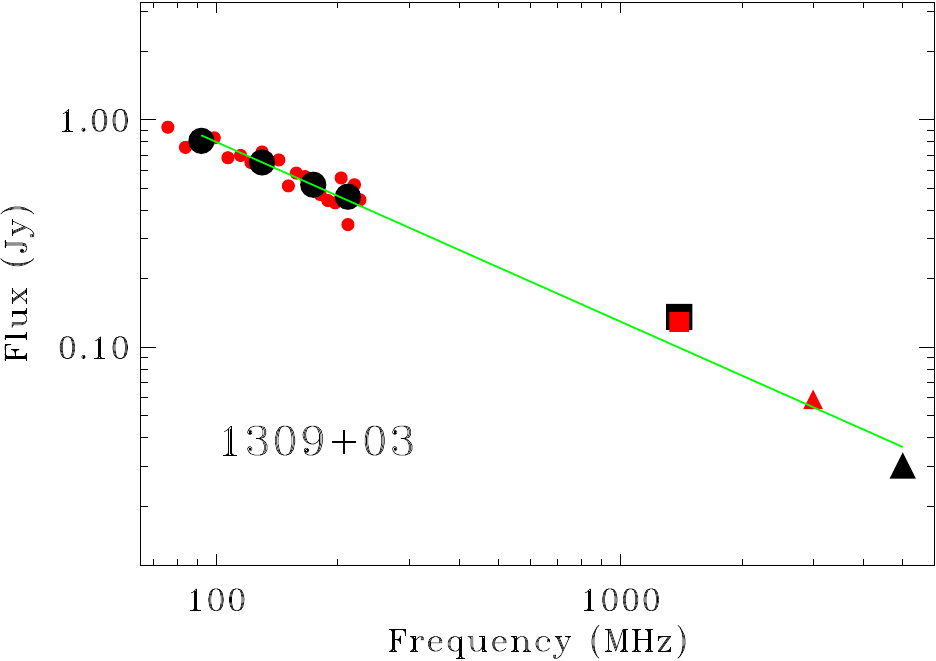}
    \includegraphics[width=0.30\textwidth]{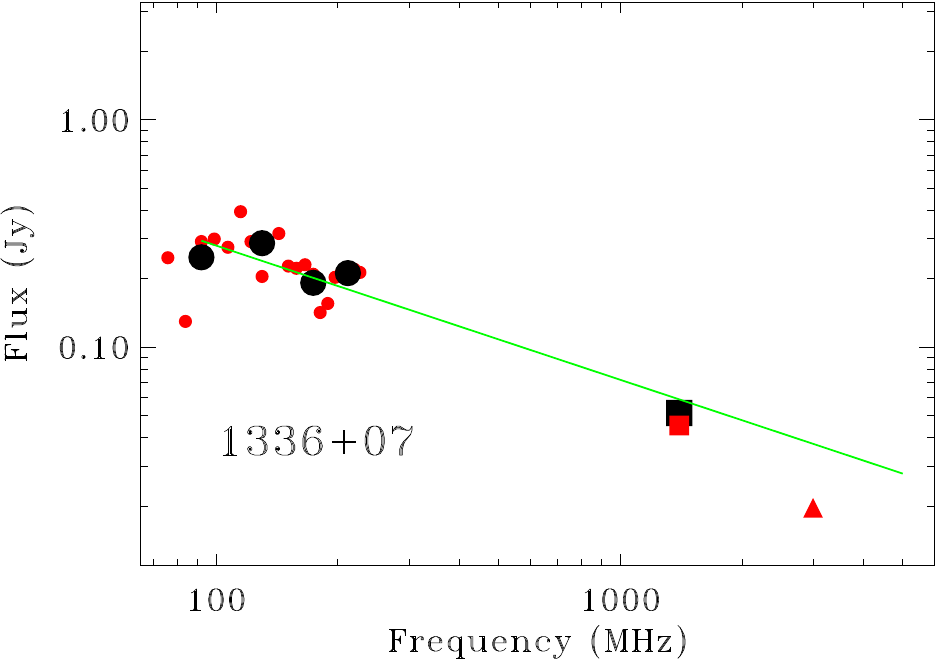}
    \includegraphics[width=0.30\textwidth]{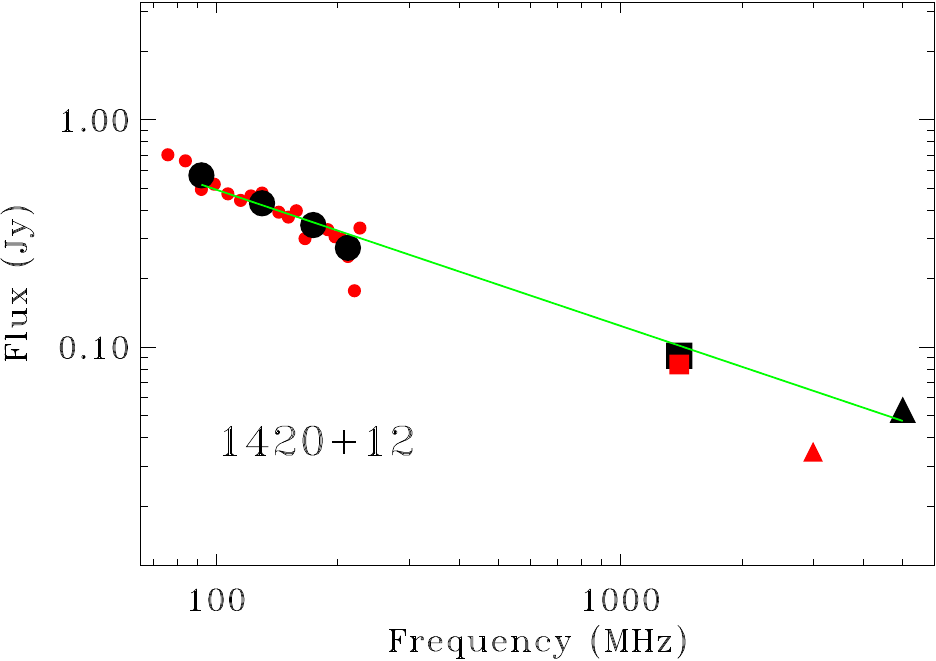}
    \includegraphics[width=0.30\textwidth]{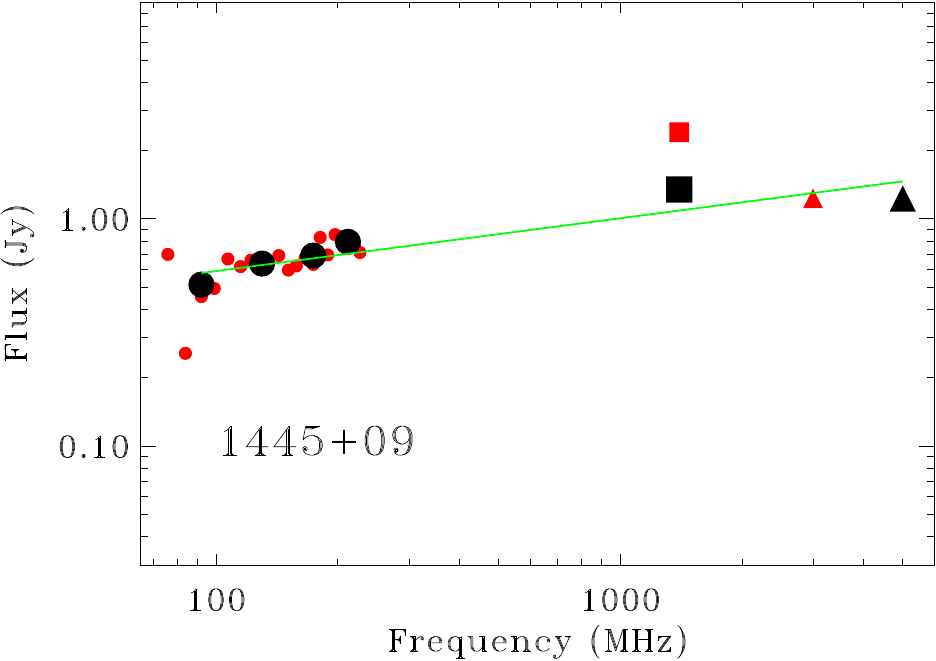}
    \includegraphics[width=0.30\textwidth]{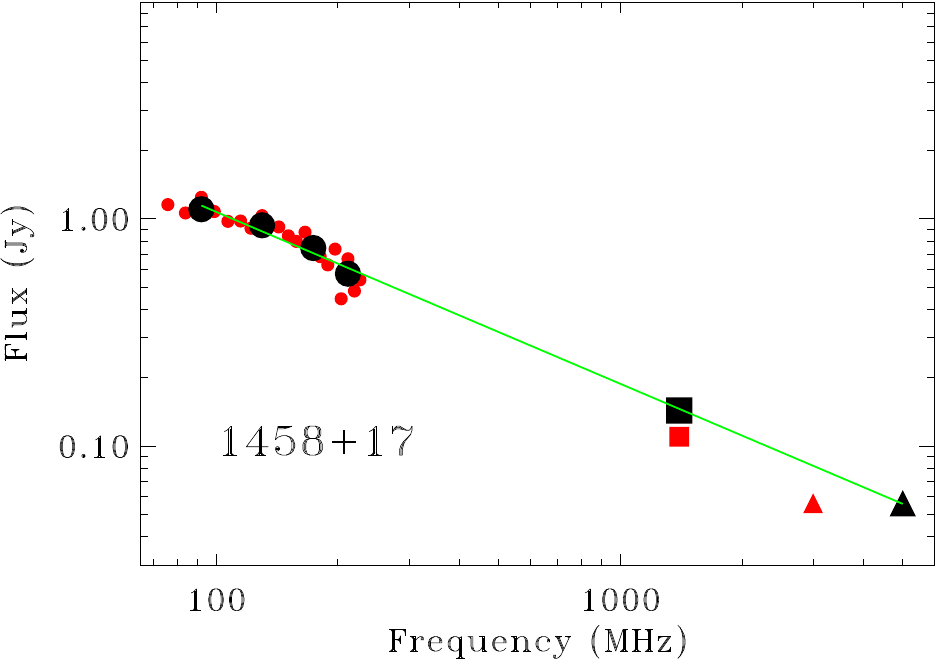}
    \includegraphics[width=0.30\textwidth]{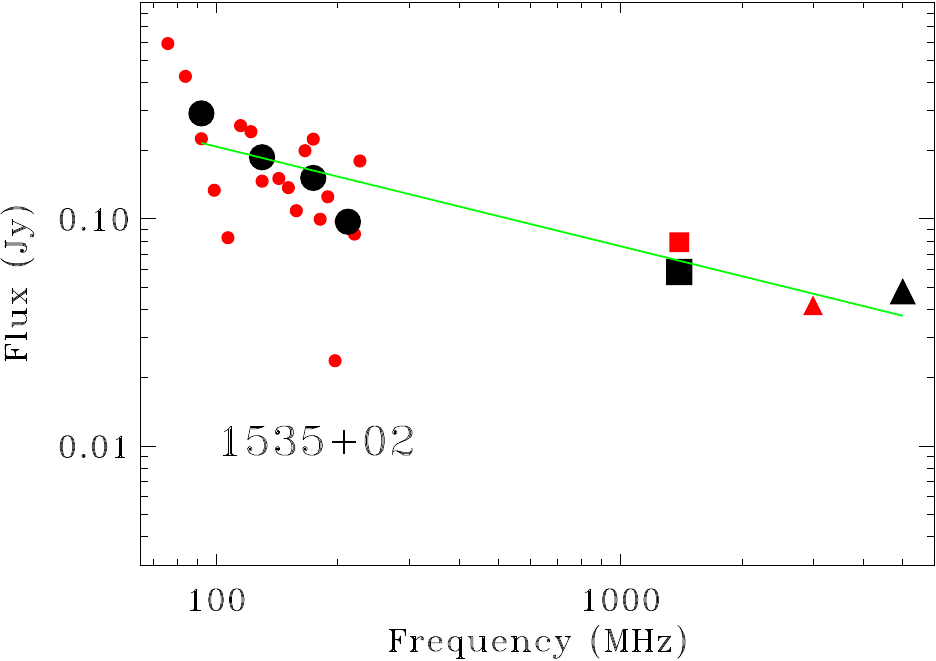}
    \includegraphics[width=0.30\textwidth]{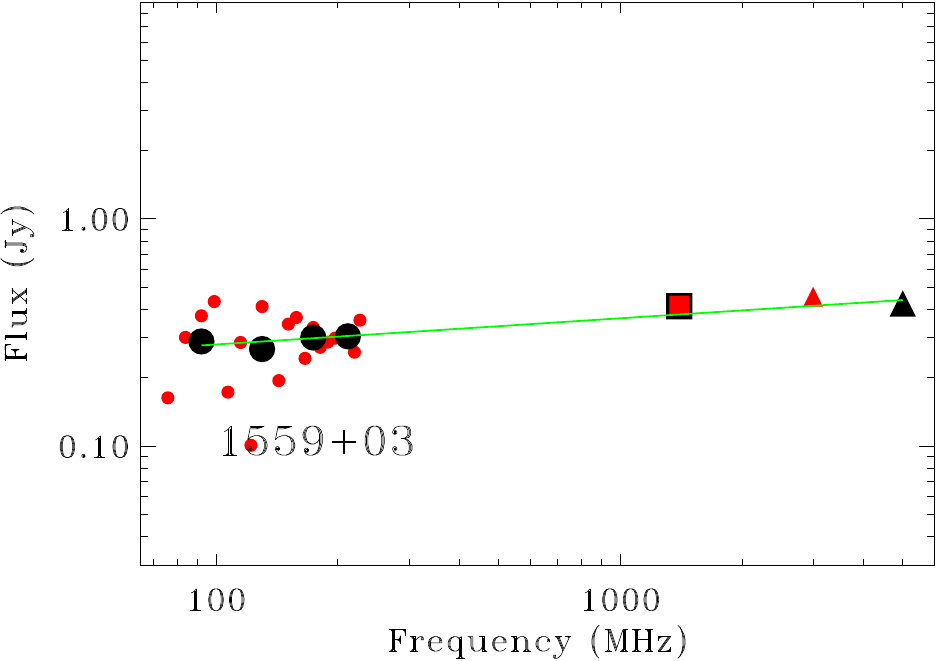}
    \includegraphics[width=0.30\textwidth]{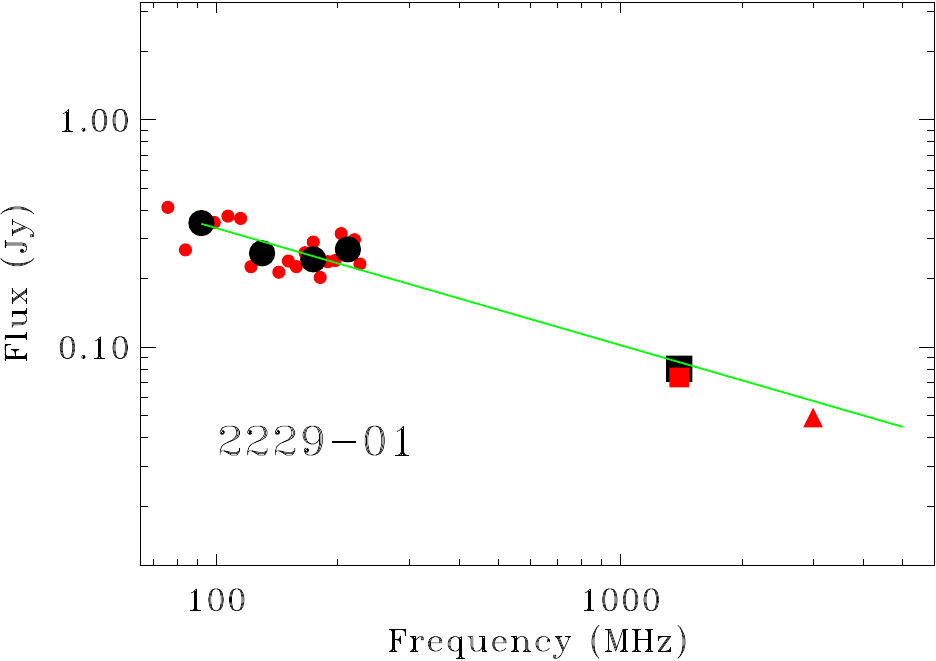}
    \caption{(continued) }                                   
    \label{spectra2}                                          
\end{figure*}                                               
\end{document}